\documentclass[fontsize=11pt,twoside=semi,numbers=endperiod]{scrartcl}
\KOMAoptions{headinclude=true,footinclude=false,index=toc}
\usepackage[paperwidth=150mm,paperheight=179.35mm,width=110mm,totalheight=165.35mm,includehead,top=4mm,inner=20mm]{geometry}

\usepackage{amsmath}
\usepackage{amssymb}

\usepackage{lmodern} 

\usepackage[ansinew]{inputenc}
\usepackage[T1]{fontenc}
\usepackage[english,ngerman]{babel}
\usepackage{textcomp} 
\usepackage[pdftex]{color,graphicx} %
\usepackage[english]{varioref} 
\usepackage{bbm}
\usepackage{cite} 
\usepackage{ellipsis} 
\usepackage{microtype} 
\usepackage[rel]{overpic} 
\usepackage{scrpage2}
\usepackage[%
pdftex,
colorlinks,citecolor=blue,filecolor=blue,linkcolor=blue,urlcolor=blue,%
bookmarksnumbered=true,
pdfpagemode=UseNone,
pdffitwindow=true,%
pdfstartview={},
pdfpagelayout=SinglePage,
plainpages=false,
pdfpagelabels=true,
]
{hyperref}
\hypersetup{%
    pdfauthor={Astrophysical Institute Neunhof},
	pdfsubject={Wheeler-Feynman Absorber Theory},
	pdfkeywords={Wheeler-Feynman Absorber Theory},
	pdftitle={Some Remarks on Wheeler-Feynman Absorber Theory}}

\setkomafont{disposition}{\rmfamily} 
\addtokomafont{section}{\large\bfseries} 
\addtokomafont{subsection}{\normalsize \mdseries \itshape} 
\addtokomafont{pagenumber}{\LARGE}
\addtokomafont{caption}{\small}

\deffootnote[1em]{1em}{1em}{\textsuperscript{\thefootnotemark}\ }

\newenvironment{ggitemize}{\vspace{0.2\baselineskip} \begin{addmargin}[1.2em]{0em}}{\vspace{-0.8\baselineskip} \end{addmargin}}
\newcommand{\ggitem}[2][$\ast $]{\makebox[0em][r]{#1\;}{#2}\newline }%
\newcommand{\bz}[1]{\nolinebreak\hspace{0em}\nolinebreak{}#1\hspace{0em}}

\newcommand{\al}{\iflanguage{english}{``\nolinebreak\hspace{0em}}{\glqq\nolinebreak\hspace{0.1em}}\nolinebreak}
\newcommand{\ar}{\nolinebreak\iflanguage{english}{\hspace{0em}\nolinebreak{}''\ }{\hspace{-0.45em}\nolinebreak\grqq\thickspace}}
\newcommand{\arp}{\nolinebreak\iflanguage{english}{\hspace{0em}\nolinebreak{}''}{\grqq}}
\newcommand{\ggie}{\mbox{i.\,e.\ }}
\newcommand{\ggeg}{\mbox{e.\,g.\ }}
\newcommand{\ggresp}{\mbox{resp.\ }}
\newcommand{\ARZ}{\nolinebreak\mbox{\hspace{.08em}!\ }}
\newcommand{\ARZp}{\nolinebreak\mbox{\hspace{.08em}!}}

\newcommand{\ggstackrel}[2][0]{\hspace{0.36em}\hspace{#1em}&=\hspace{-#1em}\hspace{-1.16em}\stackrel{#2}{\phantom{=}}} 
\newcommand{\inte }{\int \limits }

\newcommand{\difq}[2]{\ensuremath{\frac{\mathrm{d}\hspace{0.1em}#1}{\mathrm{d}\hspace{0.1em}#2}}} 
\DeclareMathOperator{\dif }{d\hspace{-0.12em}}

\definecolor{lila}{rgb}{.7,0,.8}
\definecolor{dklgrn}{rgb}{0,0.8,0}
\definecolor{hellblau}{rgb}{0.4,0.5,1}
\definecolor{ocker}{rgb}{1,0.5,0}

\clearscrheadfoot 
\rohead[\vspace{-2mm}\par\pagemark ]{\vspace{-2mm}\par\pagemark}
\lehead[\vspace{-2mm}\par\pagemark ]{\vspace{-2mm}\par\pagemark}
\rehead{ }
\lohead{ }

\begin{document}
\selectlanguage{english}
\raggedbottom  
\thispagestyle{empty}
\vspace*{-4mm}\enlargethispage{.5\baselineskip} 
\centerline{\bfseries\Large Remarks on}
\vspace*{.5mm}
\centerline{\bfseries\Large Wheeler-Feynman Absorber Theory}
\vspace*{2mm}
\centerline{{\bfseries Gerold Gr\"{u}ndler}\,\footnote{e-mail:\ gerold.gruendler@astrophys-neunhof.de}}
\vspace*{1mm}
\centerline{\small Astrophysical Institute Neunhof, N\"{u}rnberg, Germany} 
\vspace*{3mm}
\noindent\parbox{\textwidth}{The derivation of absorber\bz{-}theory is outlined in very detail. Absorber theory is based on classical action\bz{-}at\bz{-}a\bz{-}distance electrodynamics, but it deviates from that theory at a crucial point. It is shown that (a) absorber theory cannot achieve any of it's essential results without this deviation, and that (b) this deviation restricts the application range of absorber theory to stationary radiation processes. Furthermore an error which crept into Wheeler's and Feynman's interpretation of their equation (19) is pointed out. These shortcomings can probably be eliminated by a quantum\bz{-}theoretical formulation of absorber theory.} 
\section*{Contents}
\hyperlink{ta:strahlung}{1.\hspace{0.7em}Retarded and advanced fields \dotfill\ \pageref{absch:strahlung}}\\
\vspace{-0.7\baselineskip} \\
\hyperlink{ta:rueckwirkung}{2.\hspace{0.7em}Radiation back-reaction \dotfill\ \pageref{absch:rueckwirkung}}\\
\vspace{-0.7\baselineskip} \\
\hyperlink{ta:wheeler}{3.\hspace{0.7em}Wheeler-Feynman absorber theory \dotfill\ \pageref{absch:wheeler}}\\
\vspace{-0.7\baselineskip} \\
\hyperlink{ta:deriv4}{4.\hspace{0.7em}Derivation IV \dotfill\ \pageref{absch:deriv4}}\\
\vspace{-0.7\baselineskip} \\
\hyperlink{ta:problem1}{5.\hspace{0.7em}Applying the reasonable assumption (15b) \dotfill\ \pageref{absch:problem1}}\\
\vspace{-0.7\baselineskip} \\
\hyperlink{ta:why}{6.\hspace{0.7em}The perplexing postulate (15c) \dotfill\ \pageref{absch:why}}\\
\vspace{-0.7\baselineskip} \\
\hyperlink{ta:error19}{7.\hspace{0.7em}Correcting the interpretation of (42a) \dotfill\ \pageref{absch:error19}}\\
\vspace{-0.7\baselineskip} \\
\hyperlink{ta:altern}{8.\hspace{0.7em}Quantized absorber theory? \dotfill\ \pageref{absch:altern}}\\
\vspace{-0.7\baselineskip} \\
\hyperlink{ta:zusfass}{9.\hspace{0.7em}Conclusions \dotfill\ \pageref{absch:zusfass}}\\
\vspace{-0.7\baselineskip} \\
\hyperlink{ta:literatur}{$\phantom{9.}$\hspace{0.7em}References \dotfill\ \pageref{absch:literatur}} 

\section{Retarded and advanced fields}\label{absch:strahlung} 
Let \raisebox{4\baselineskip}[0pt][0pt]{\hypertarget{ta:strahlung}{}}a charged point\bz{-}particle $q$ be at time $t_s$ at position $\boldsymbol{r}$ in three\bz{-}dimensional space. Let it's velocity (which also may be zero) be $\boldsymbol{v}$, and it's acceleration (which also may be zero) be $\boldsymbol{\dot{v}}$. The electric and magnetic fields generated by the point charge are measured at space\bz{-}time point $x=(t,\boldsymbol{x})$. The computation of the fields is demonstrated in very detail in \cite{apin:se90115}. Using the three\bz{-}dimensional vectors 
\begin{align}
\boldsymbol{R}\equiv\boldsymbol{x}-\boldsymbol{r}\quad ,\quad 
R\equiv |\boldsymbol{R}|\quad ,\quad\boldsymbol{n\equiv}\frac{\boldsymbol{R}}{R}\label{ksghsdfg}
\end{align}
the results are 
\begin{subequations}\label{lksjngfdkf}\begin{align}
\boldsymbol{E}_s(x)&=\frac{1}{4\pi\epsilon _0}\, \frac{q(\boldsymbol{n} \mp\boldsymbol{v}/c)}{R^2\gamma ^2(1\mp\boldsymbol{n\cdot v}/c)^3}\bigg| _{t _s}+\notag\\ 
&\quad +\frac{1}{4\pi\epsilon _0}\, \frac{q\,\boldsymbol{n\times}\Big( (\boldsymbol{n}\mp\boldsymbol{v}/c)\boldsymbol{\times\dot{v}}\Big)}{Rc^2(1\mp\boldsymbol{n\cdot v}/c)^3}\,\bigg| _{t _s}\label{lksjngfdkfb}\displaybreak[1]\\ 
\boldsymbol{B}_s(x)&=-\frac{\mu _0c}{4\pi}\, \frac{q\, (\boldsymbol{n\times v}/c)}{R^2\gamma ^2(1\mp\boldsymbol{n\cdot v}/c)^3}\,\bigg| _{t _s}\, -\notag\\  
&\quad -\frac{\mu _0c}{4\pi}\, \frac{q}{Rc^2(1\mp\boldsymbol{n\cdot v}/c)^3}\,\cdot\bigg[ (\boldsymbol{n\times v}/c)(\boldsymbol{n\cdot\dot{v}})\, +\notag\\ 
&\hspace{3em}+(\boldsymbol{n\times\dot{v}})(\pm 1-\boldsymbol{n\cdot v}/c )\bigg] _{\tau _s}\label{lksjngfdkfd}\displaybreak[1]\\
&\text{with}\quad t=t _s\pm R/c=t _r+R/c=t _a-R/c\ .\notag    
\end{align}\end{subequations} 
The index $_s$ may assume the values $_r=\text{retarded}$ or $_a=\text{advanced}$.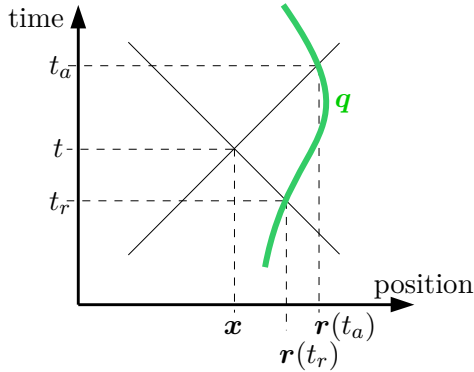
\begin{figure}[t!]\centering\begin{overpic}[trim=0 6 0 0]{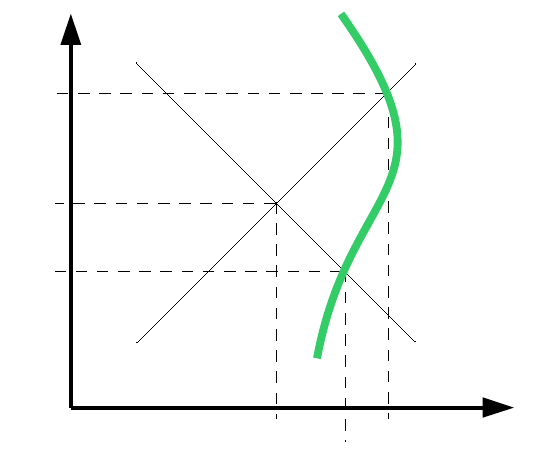}
\put(75.5,56){\textcolor{dklgrn}{$\boldsymbol{q}$}}
\put(48.5,1){$\boldsymbol{x}$}
\put(62,-6){$\boldsymbol{r}(t_r)$}
\put(70.5,1){$\boldsymbol{r}(t_a)$}
\put(85,11){position}
\put(6,64){$t_a$}
\put(7,44){$t$}
\put(6,31){$t_r$}
\put(-4,75.5){time}
\end{overpic}\caption{Worldline of the charge $q$}\label{fig:lichtkegel}\end{figure} Double signs $\pm $ or $\mp$ shall be interpreted as \mbox{\,{\tiny advanced}\hspace{-2.7em}\raisebox{2mm}{{\tiny retarded}}\hspace{.3em}}, \ggie the upper sign always holds for the retarded fields, the lower sign for the advanced fields. The symbol $| _{t _s}$ is indicating that $\boldsymbol{n}(t_s)$, $\boldsymbol{v}(t_s)$, and $R(t_s)$ shall be inserted. Note that  $R(t_r)=|\boldsymbol{x}-\boldsymbol{r}(t_r)|\neq R(t_a)=|\boldsymbol{x}-\boldsymbol{r}(t_a)|$. SI\bz{-}units\!\cite{SI:units} are used throughout this article. 

In figure~\ref{fig:lichtkegel} \vpageref[\negmedspace ]{fig:lichtkegel} the worldline of the point\bz{-}charge $q$ is sketched in green, and the lightcone of the observation point $(t,\boldsymbol{x})$ is indicated. Retarded fields, which are observed at time $t$ at position $\boldsymbol{x}$, have been emitted by the charge $q$ at the previous time $t_r=t-R/c$. Advanced fields, which are observed at time $t$ at position $\boldsymbol{x}$, will be emitted by the charge $q$ only in future, namely at the later time $t_a=t+R/c$, and propagate towards $x$ backwards through time. 

Advanced fields seemingly are violating the \al principle of causality\arp . Therefore many authors regard them as \al unphysical\arp , and skip them from the outset. If a flash of light is emitted in direction of a mirror, then the reflected flash is always observed after, but never before the emission. Still there is an alternative to Maxwell's electrodynamics, namely action\bz{-}at\bz{-}a\bz{-}distance electrodynamics, which has been proposed and worked out in particular by Gau\ss, Schwarzschild, and Frenkel (see\!\cite{Wheeler:actiondist,Hoylenarl:actatdist} and the references cited in these articles). This theory claims to be able to describe all observable electrodynamic phenomena as good as Maxwell's theory. In action\bz{-}at\bz{-}a\bz{-}distance theory, fields are no self\bz{-}contained physical objects. Instead this theory only knows charged particles, which interact due to action\bz{-}at\bz{-}a\bz{-}distance forces. Fields are considered to be only computing aids on the theorist's paper. 

Both the retarded and the advanced computing aids, which we will continue to name fields, are indispensable for action\bz{-}at\bz{-}a\bz{-}distance electrodynamics. Wheeler and Feynman devised absorber theory within the framework of action\bz{-}at\bz{-}a\bz{-}distance electrodynamics. Therefore we necessarily must keep the advanced fields. 

The relation 
\begin{align}
\pm\boldsymbol{n}(t _s)\boldsymbol{\times E}_s(x)=c\boldsymbol{B}_s(x)\label{kshgsgvsd} 
\end{align}
holds for the electric and magnetic fields \eqref{lksjngfdkf}, see \cite{apin:se90115}. Thus the magnetic field can be computed easily, once the electric field is known. Therefore in the sequel we will restrict most computations to the electric fields. Furthermore we emphasize, that the fields \eqref{lksjngfdkf} are relativistically invariant, even though this is not obvious in the three\bz{-}dimensional notation. 

The fields \eqref{lksjngfdkf}, and in particular the reasons and consequences of their various signs, are discussed thoroughly in \cite{apin:se90115}. In the following investigation, we will consider only the second terms respectively, which are proportional to $R^{-1}$, because in the Poynting\bz{-}vector 
\begin{align}
\boldsymbol{S}=\frac{1}{\mu _0}\,\boldsymbol{E_s\times B_s}\stackrel{\eqref{kshgsgvsd}}{=}\pm\frac{1}{c\mu _0}\,\boldsymbol{E_s\times}\! (\boldsymbol{n\times E_s})\ ,\label{ksdghnsdgsa} 
\end{align}
the product of these two terms is the only one $\sim\! R^{-2}$. All products containing one ore two of the other terms are $\sim\! R^{-3}$ or $\sim\! R^{-4}$, and therefore are significant only nearby the source $q$. For this reason, the terms $\sim\! R^{-1}$ are called radiation fields by many authors. Note, that radiation fields are proportional to $\boldsymbol{\dot{v}}$. They vanish if the source is at rest or in uniform motion. Note furthermore the negative sign of the advanced Poynting\bz{-}vector: Advanced fields propagate opposite to the \al usual\ar  direction of time from the future to the past. The retarded radiation fields form spherical waves, which are emitted by $q$ at time $t_r$, and spread off from the source $q$. The advanced fields form spherical waves, which are approaching in the past from infinite distance, are contracting closer and closer around the position of the source $q$, and collapse onto $q$ at time $t_a$. Thus the stream of advanced energy is directed towards the source, while the stream of retarded energy is directed away from the source. 

\begin{figure}[b!]\centering\begin{overpic}[trim=0 0 0 -4]{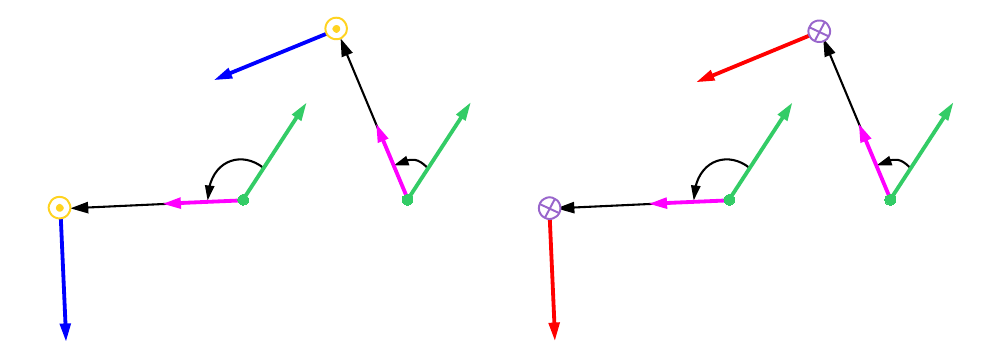}
\put(97,21){$\textcolor{dklgrn}{\boldsymbol{\dot{v}}}$}
\put(81,21){$\textcolor{dklgrn}{\boldsymbol{\dot{v}}}$}
\put(48,21){$\textcolor{dklgrn}{\boldsymbol{\dot{v}}}$}
\put(31,21){$\textcolor{dklgrn}{\boldsymbol{\dot{v}}}$}
\put(90.5,19.5){$\vartheta $}
\put(72,19.5){$\vartheta $}
\put(41.5,19.5){$\vartheta $}
\put(23,19.5){$\vartheta $}
\put(92,13.5){$\textcolor{dklgrn}{q}$}
\put(76,13.5){$\textcolor{dklgrn}{q}$}
\put(42.5,13.5){$\textcolor{dklgrn}{q}$}
\put(26,13.5){$\textcolor{dklgrn}{q}$}
\put(89,11.5){$\boldsymbol{r}$}
\put(73,11.5){$\boldsymbol{r}$}
\put(40,11.5){$\boldsymbol{r}$}
\put(23,11.5){$\boldsymbol{r}$}
\put(85.5,17.5){$\textcolor{magenta}{\boldsymbol{n}}$}
\put(69,11.5){$\textcolor{magenta}{\boldsymbol{n}}$}
\put(36.5,17.5){$\textcolor{magenta}{\boldsymbol{n}}$}
\put(19,11.5){$\textcolor{magenta}{\boldsymbol{n}}$}
\put(87,25.5){$\boldsymbol{R}$}
\put(61,10.5){$\boldsymbol{R}$}
\put(37.5,25.5){$\boldsymbol{R}$}
\put(11,10.5){$\boldsymbol{R}$}
\put(79.5,32.5){$\boldsymbol{x}$}
\put(52,11.5){$\boldsymbol{x}$}
\put(30.5,32.5){$\boldsymbol{x}$}
\put(2,11.5){$\boldsymbol{x}$}
\put(83,34){$\textcolor{lila}{\boldsymbol{B}_a}$}
\put(54.5,15.5){$\textcolor{lila}{\boldsymbol{B}_a}$}
\put(34,34){$\textcolor{ocker}{\boldsymbol{B}_r}$}
\put(4,15.5){$\textcolor{ocker}{\boldsymbol{B}_r}$}
\put(70.5,30){$\textcolor{red}{\boldsymbol{E}_a}$}
\put(57.5,3){$\textcolor{red}{\boldsymbol{E}_a}$}
\put(21,30){$\textcolor{blue}{\boldsymbol{E}_r}$}
\put(7.5,3){$\textcolor{blue}{\boldsymbol{E}_r}$}
\end{overpic}\caption{The radiation fields of the source $q$}\label{fig:geschwubeschl}\end{figure}
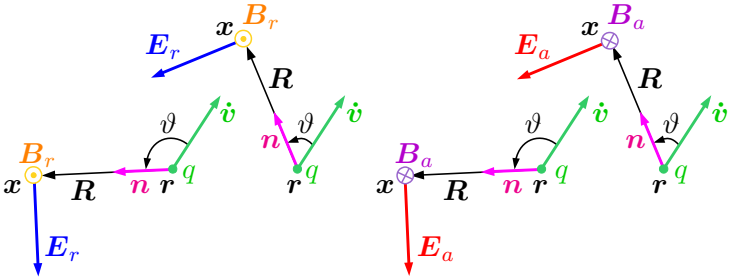 All formulas indicated so far are valid for arbitrary velocities of the source $q$, including relativistic velocities $(v/c\lessapprox 1)$. For the following investigations the non\bz{-}relativistic approximations $(v/c\ll 1)$ of the radiation fields, which can easily be extracted from \eqref{lksjngfdkf} and \eqref{kshgsgvsd}, will be sufficient: 
\begin{subequations}\label{klsngmsgsd}\begin{align} 
\boldsymbol{E}_s(x)&=\frac{1}{4\pi\epsilon _0}\, \frac{q\,\boldsymbol{n\times}\! (\boldsymbol{n\times\dot{v}})}{Rc^2}\Big| _{t _s} \label{oshgkdsdgsda}\displaybreak[1]\\
\Big|\boldsymbol{E}_s(x)\Big| &=\frac{1}{4\pi\epsilon _0}\, \frac{q\,\sin (\boldsymbol{n},\boldsymbol{\dot{v}})}{Rc^2}\Big| _{t _s}\displaybreak[1]\\
c\boldsymbol{B}_s(x)&=\pm\boldsymbol{n}(t _r)\boldsymbol{\times E}_r^{(s)}\! (x)\label{oshgkdsdgsdb}\\ 
\text{if }v/c\ll 1\quad &,\quad\text{with }t=t _s\pm R/c=t _r+R/c=t _a-R/c\notag   
\end{align}\end{subequations} 
In the non\bz{-}relativistic case, $\boldsymbol{E}_s$ lies in the plane defined by $\boldsymbol{n}$ and $\boldsymbol{\dot{v}}$ (see \mbox{fig.\,\ref{fig:geschwubeschl}}\vpageref[\negmedspace ]{fig:geschwubeschl}), is perpendicular to $\boldsymbol{n}$, and the angle $(\boldsymbol{E}_s,\boldsymbol{\dot{v}})$  is always $\geq\pi /2$ in case $q>0$. The retarded vector $\boldsymbol{B}_r$ is pointing vertically up from the drawing plane, while the advanced vector $\boldsymbol{B}_a$ is pointing vertically down into the drawing plane. 

\section{Radiation back-reaction}\label{absch:rueckwirkung} 
If \raisebox{4\baselineskip}[0pt][0pt]{\hypertarget{ta:rueckwirkung}{}}an uncharged body is accelerated by an external force, then an inertial force of same strength but opposite direction is opposing the acceleration. If the accelerated body is electrically charged, then an additional force is opposing the external accelerating force. This additional force is called radiation back\bz{-}reaction. The external accelerating force must do additional work, because the accelerated charged body is immediately radiating a part of the aquired energy in form of electromagnetic waves. 

Abraham and Lorentz computed the (non\bz{-}relativistic) radiation back\bz{-}reaction force: 
\begin{align}
\boldsymbol{F}_{\text{rad}}=\frac{2q^2\boldsymbol{\ddot{v}}}{3c^34\pi\epsilon _0}\quad\text{if }v\ll c\label{ksdgvsgc} 
\end{align} 
They approached this result by two different methods: One method was based on considerations with regard to energy conservation. Their second computation, in which they considered in particular an accelerated electron, was based on a classical model of the electron: They assumed the electron to be a sphere with radius of about \mbox{$3\cdot 10^{-15}\text{m}$}, and they demonstrated that the retarded force, which the electron is exerting onto itself by means of the fields which it is radiating, just has the value \eqref{ksdgvsgc}. Both derivations are presented in very detail in \cite{apin:se90115}. 

By today, however, it is known from experimental investigations, that the radius of the electron, provided it should be finite, must be significantly smaller than \mbox{$10^{-18}\text{m}$}. Therefore the electron model of Abraham and Lorentz can not be correct. In contrast, the alternative derivation of \eqref{ksdgvsgc}, which they found on grounds of energetic considerations, is still valid by today (though not free of severe problems, as discussed in \cite[sect.\,3]{apin:se90115}). 

It was one of the central motivations for the absorber theory of Wheeler and Feynman\!\cite{Wheeler:absorber}, to derive an explanation for \eqref{ksdgvsgc} under the assumption, that electrons (and other elementary particles) are point\bz{-}particles with zero radius. According to absorber theory, the radiation back\bz{-}reaction onto an accelerated charge $q$ is caused by the advanced electromagnetic fields radiated by those particles which absorb the energy radiated by $q$. In the next section we will provide an in\bz{-}depth outline of absorber theory. 

\section{Wheeler-Feynman absorber theory}\label{absch:wheeler}
A \raisebox{4\baselineskip}[0pt][0pt]{\hypertarget{ta:wheeler}{}}particle with charge $q$ is accelerated by $\boldsymbol{\dot{v}}$ at the retarded time $t_r=t-R/c$ at position $\boldsymbol{r}$. The accelerated particle is emitting the retarded fields indicated in \eqref{lksjngfdkf}. (It will radiate advanced fields as well; but we ignore them for the moment being.) We restrict the investigation to the non\bz{-}relativistic case $v\ll c$, in which the radiation fields emitted by the particle assume the simple form \eqref{klsngmsgsd}. 

Let a particle with charge $q_k$, mass $m_k$, and velocity $v_k\ll c$ be at space\bz{-}time point $x_k$. The retarded fields $\boldsymbol{E}_r$ and $\boldsymbol{B}_r$ exert onto $q_k$ in the non\bz{-}relativistic case the Lorentz force  
\begin{align}
m_k\,\difq{\boldsymbol{v}_k}{t}=q_k(\boldsymbol{E}_r+\boldsymbol{v}_k\boldsymbol{\times B}_r)\ .\label{xdsfdgjbgb}  
\end{align} 
If the velocities of the particles are not only non\bz{-}relativistic 
\begin{align} 
v\ll c\qquad ,\qquad v_k\ll c \ ,\label{lsjngdksgd} 
\end{align} 
but if in addition 
\begin{align}
|\boldsymbol{v}_k\boldsymbol{\times B}_r|\ll |\boldsymbol{E}_r|\label{oskgesfger}
\end{align} 
holds, then the acceleration of the particle $q_k$ at time $t$ and at position $\boldsymbol{x}_k$ is  
\begin{align}
\boldsymbol{\dot{v}}_k\approx\frac{q_k\boldsymbol{E}_r(x_k)}{m_k}\stackrel{\eqref{klsngmsgsd}}{=}\frac{1}{4\pi\epsilon _0}\, \frac{q_kq\boldsymbol{n\times}\! (\boldsymbol{n\times\dot{v}})}{m_kRc^2}\Big| _{t_r}\ .\label{oiskgjhs}
\end{align} 

Now Wheeler and Feynman assume, that the accelerated point\bz{-}charge $q_k$ will radiate for it's part not only a retarded field, but also an advanced field of same strength. Thereby the retarded and advanced fields shall have only half the amplitude of the fields indicated in \eqref{klsngmsgsd}. The factors 1/2 will later be justified by the results, which follow from this ansatz. 

\begin{figure}[b!]\centering\begin{overpic}[trim=0 0 0 -4]{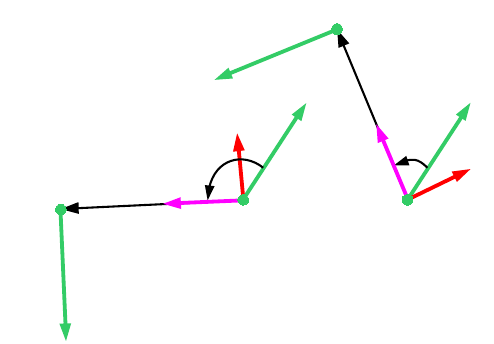}
\put(96,42){$\textcolor{dklgrn}{\boldsymbol{\dot{v}}}$}
\put(62,42){$\textcolor{dklgrn}{\boldsymbol{\dot{v}}}$}
\put(83,39){$\vartheta $}
\put(38,33){$\vartheta $}
\put(81,23.5){$\textcolor{dklgrn}{q}$}
\put(52,27){$\textcolor{dklgrn}{q}$}
\put(76,27){$\boldsymbol{r}$}
\put(46,23){$\boldsymbol{r}$}
\put(72.5,35){$\textcolor{magenta}{\boldsymbol{n}}$}
\put(38,23){$\textcolor{magenta}{\boldsymbol{n}}$}
\put(75,51){$\boldsymbol{R}$}
\put(22,21){$\boldsymbol{R}$}
\put(61,67){$\boldsymbol{x}_k$}
\put(3,24){$\boldsymbol{x}_k$}
\put(71,64){$\textcolor{dklgrn}{q_k}$}
\put(8,31){$\textcolor{dklgrn}{q_k}$}
\put(42,60){$\textcolor{dklgrn}{\boldsymbol{\dot{v}}_k}$}
\put(15,6){$\textcolor{dklgrn}{\boldsymbol{\dot{v}}_k}$}
\put(90,26){$\textcolor{red}{\boldsymbol{E}_a}$}
\put(43,43){$\textcolor{red}{\boldsymbol{E}_a}$}
\end{overpic}\caption{The advanced radiation fields of the accelerated charge $q_k$}\label{fig:wheelfeyn2}\end{figure}
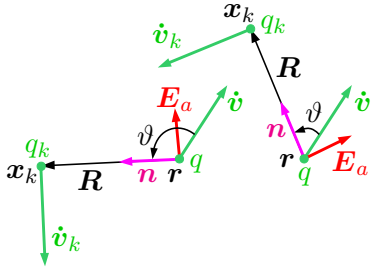

The advanced fields radiated by $q_k$ move backwards through time, and arrive at the position $\boldsymbol{r}$ of the primary source just at the time $t_r$, at which this source is radiating the retarded fields \eqref{klsngmsgsd}: 
\begin{align}
\boldsymbol{E}_a(t_r,\boldsymbol{r})&\stackrel{\eqref{klsngmsgsd}}{=}\frac{1}{2}\,\frac{1}{4\pi\epsilon _0}\, \frac{q\,\boldsymbol{n}_k\!\boldsymbol{\times}\! (\boldsymbol{n}_k\boldsymbol{\times\dot{v}}_k)}{Rc^2}\Big| _{t}\label{klsahgdsgd}\\ 
&\text{with }t_r=t-R/c\notag   
\end{align}
Obviously $\boldsymbol{n}_k=-\boldsymbol{n}$ holds for $\boldsymbol{n}_k$ in equation \eqref{klsahgdsgd} and $\boldsymbol{n}$ in equation \eqref{oiskgjhs}. Therefore we will write $-\boldsymbol{n}$ instead of $\boldsymbol{n}_k$. Furthermore we stipulate that --- as already exercised in \eqref{klsahgdsgd} --- $t_r$ shall always denote the time, at which $q$ radiates the retarded field. Thus $\boldsymbol{n}$ is directed from $\boldsymbol{r}$ (the position of $q$ at time $t_r$) towards $\boldsymbol{x}_k$ (the position of $q_k$ at time $t$). The advanced fields radiated by $q_k$ are at the position $\boldsymbol{r}$ of the primary source $q$ equal to 
\begin{subequations}\label{oasaghdews}\begin{align}
\boldsymbol{E}_a(t_r,\boldsymbol{r})&=\frac{1}{4\pi\epsilon _0}\, \frac{q_k\boldsymbol{n\times}(\boldsymbol{n\times\dot{v}}_k)}{2Rc^2}\Big| _{t}\\ 
\ggstackrel{\eqref{oiskgjhs}}\frac{1}{(4\pi\epsilon _0)^2}\, \frac{q_k^2q\boldsymbol{n\times}\! (\boldsymbol{n\times}(\boldsymbol{n\times}(\boldsymbol{n\times\dot{v}})))}{2m_kR^2c^4}\Big| _{t_r}\displaybreak[1]\\ 
\Big|\boldsymbol{E}_a(t_r,\boldsymbol{r})\Big| &=\frac{1}{4\pi\epsilon _0}\, \frac{|q_k\boldsymbol{\dot{v}}_k|}{2Rc^2}\Big| _{t}\quad\text{because of }|\sin (\boldsymbol{n},\boldsymbol{\dot{v}}_k)|=1\\ 
&=\frac{1}{(4\pi\epsilon _0)^2}\, \frac{q_k^2\, |q\dot{v}|\,\sin (\boldsymbol{n},\boldsymbol{\dot{v}})}{2m_kR^2c^4}\Big| _{t_r}\displaybreak[1]\\ 
\boldsymbol{B}_a(t_r,\boldsymbol{r})&=+\frac{1}{c}\, \boldsymbol{n\times E}_a(t_r,\boldsymbol{r})\ .  
\end{align}\end{subequations} 
In fig.\,\ref{fig:wheelfeyn2} \vpageref[\negmedspace ]{fig:wheelfeyn2} (which should be compared to fig.\,\ref{fig:geschwubeschl}\ARZp ) 
the advanced fields radiated by $q_k$ are sketched for the non\bz{-}relativistic case under the assumption $q_k>0$ and $q>0$. 

If the radiation emitted by $q$ is absorbed by many particles $q_k$ at different distances from $q$, and if the particles $q_k$ thereupon for their parts radiate advanced fields, then these fields will superpose to a total field at the position $\boldsymbol{r}$ of $q$. It is the central idea of absorber theory, that the retarded field radiated by $q$ will sooner or later be completely absorbed by other particles $q_k$. Each absorber particle $q_k$ thereupon emits retarded fields, and advanced fields of same strength. The superposition of all the advanced fields arriving at the primary source $q$ at time $t_r$ then shall bring about the radiation\bz{-}reaction force acting upon $q$. 

From \mbox{fig.\,\ref{fig:wheelfeyn2}} however it is visible that the angle enclosed by $\boldsymbol{\dot{v}}$ and $\boldsymbol{E}_a(\boldsymbol{r})$ is $\leq\pi /2$ for any value of $\vartheta $. The force $\boldsymbol{F}=q\boldsymbol{E}_a$ exerted by the advanced fields onto the primary source $q$ thus increases the original acceleration $\boldsymbol{\dot{v}}$, and it seems impossible that the advanced fields radiated by all absorber particles $q_k$ could add up to a back\bz{-}reaction force which is directed opposite to $\boldsymbol{\dot{v}}$. 

At this point we must bear in mind that we are dealing with time\bz{-}dependent fields. Let's consider the component $\widetilde{\boldsymbol{\dot{v}}}(\omega )$ in the Fourier\bz{-}transformation of the acceleration of $q$: 
\begin{subequations}\begin{align}
\boldsymbol{\dot{v}}(t _r,\boldsymbol{r})&=\inte _{-\infty}^{+\infty}\!\frac{\dif\omega}{2\pi}\,\widetilde{\boldsymbol{\dot{v}}}(\omega ,\boldsymbol{r})\,\exp\{ -i\omega t_r\}\label{lsdghjsava}\displaybreak[1]\\ 
\boldsymbol{\ddot{v}}&=\difq{\boldsymbol{\dot{v}}}{t}=-\inte _{-\infty}^{+\infty}\!\frac{\dif\omega}{2\pi}\, i\omega\widetilde{\boldsymbol{\dot{v}}}\,\exp\{ -i\omega t_r\}\notag\\  
&\Longrightarrow\quad \widetilde{\boldsymbol{\ddot{v}}}(\omega )=-i\omega\widetilde{\boldsymbol{\dot{v}}}\label{lsdghjsavc} 
\end{align}\end{subequations} 

\begin{figure}[!b]\centering\begin{overpic}[scale=1.8,trim=0 10 0 0]{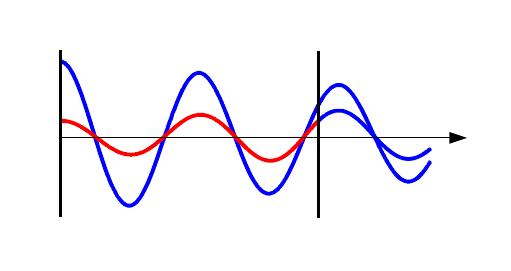}
\put(16,31){$\textcolor{blue}{\widetilde{\boldsymbol{E}}_r}$}
\put(70.5,27.5){$\textcolor{blue}{\widetilde{\boldsymbol{E}}_r}$}
\put(64.5,20){$\textcolor{blue}{\widetilde{\boldsymbol{E}}_r}$}
\put(53,20.5){$\textcolor{red}{\widetilde{\boldsymbol{E}}_a}$}
\put(13.5,15){$\textcolor{red}{\widetilde{\boldsymbol{E}}_a}$}
\put(11,0.5){$\boldsymbol{r}$}
\put(61,0.5){$\boldsymbol{x}_k$}
\put(11,38.5){$q$}
\put(61,38.5){$q_k$}
\put(87.5,21){$ct$}
\end{overpic}\caption{Phaseshift between the particles}\label{fig:phasen}\end{figure}
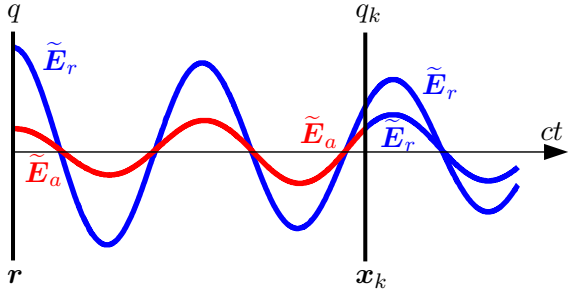 
The Fourier\bz{-}components of the various fields with frequency $\omega $ are sketched in figure~\ref{fig:phasen}\,.  At position $\boldsymbol{r}$ the accelerated charge $q$ is emitting a blue\bz{-}painted retarded field, which is in phase with $\widetilde{\boldsymbol{\dot{v}}}$. This field is accelerating at position $\boldsymbol{x}_k$ the point\bz{-}charge $q_k$ with $\widetilde{\boldsymbol{\dot{v}}}_k$. Caused by this acceleration, the charge $q_k$ is emitting for it's part a blue\bz{-}painted retarded field into the future (\ggie in the sketch towards right), and a red\bz{-}painted field of same amplitude into the past (\ggie in the sketch towards left). All field amplitudes decrease as $1/R$ with the distance $R$ from their respective sources. The energy loss of the primary retarded field due to the excitation of $q_k$ is ignored not only in the sketch, but also in the computations. We will return to this important topic immediately. 

Careful book\bz{-}keeping is necessary with regard to the phases of the various fields and accelerated particles: First the charge $q$, whose acceleration is $\boldsymbol{\dot{v}}$, radiates at position $\boldsymbol{r}$ at time $t_r$ the retarded field $\boldsymbol{E}_r$. At the moment of it's emission, this field is in phase with $\boldsymbol{\dot{v}}$: 
\begin{subequations}\label{klsjhgfdse}\begin{align}
\boldsymbol{\dot{v}}(t _r,\boldsymbol{r})\sim\boldsymbol{E}_r(t _r,\boldsymbol{r})\stackrel{\eqref{lsdghjsava}}{=}
\inte _{-\infty}^{+\infty}\!\frac{\dif\omega}{2\pi}\,\widetilde{\boldsymbol{E}}_r(\omega ,\boldsymbol{r})\,\exp\{ -i\omega t_r\} \label{olksagjsfdyasa}
\end{align}
The field propagates to the charge $q_k$ at position $\boldsymbol{x}_k$. There it arrives at time $t_k=t_r+R/c$. It's Fourier\bz{-}decomposition by then is 
\begin{align}
\boldsymbol{E}_r(t _k,\boldsymbol{x}_k)=\inte _{-\infty}^{+\infty}\!\frac{\dif\omega}{2\pi}\,\widetilde{\boldsymbol{E}}_r(\omega ,\boldsymbol{x}_k)\,\exp\{ -i\omega (\underbrace{t_r+R/c}_{t_k})\} \ . 
\end{align}
The charge $q_k$ is accelerated: 
\begin{align}
\boldsymbol{E}_r(t _k,\boldsymbol{x}_k)\!\sim\boldsymbol{\dot{v}}_k(t _k,\boldsymbol{x}_k)=\!\inte _{-\infty}^{+\infty}\!\frac{\dif\omega}{2\pi}\,\widetilde{\boldsymbol{\dot{v}}}_k(\omega ,\boldsymbol{x}_k)\,\exp\{ -i\omega (\underbrace{t_r+R/c}_{t_k})\} 
\end{align}
Caused by this acceleration, $q_k$ emits a retarded field into the future, and an advanced field into the past. According to \eqref{klsngmsgsd}, both radiated fields are in phase with $\boldsymbol{\dot{v}}_k$ at the position of $q_k$. At time $t_k-R/c=t_r$ the advanced field arrives at the position $\boldsymbol{r}$ of the charge $q$ 
\begin{align}
\boldsymbol{E}_a(t_r,\boldsymbol{r})&=\inte _{-\infty}^{+\infty}\!\frac{\dif\omega}{2\pi}\,\widetilde{\boldsymbol{E}}_a(\omega ,\boldsymbol{x})\,\exp\{ -i\omega (\underbrace{t_r+R/c}_{t_k}-R/c)\}\notag \\ 
\ggstackrel[.1]{\eqref{olksagjsfdyasa}}{\sim} \boldsymbol{\dot{v}}(t _r,\boldsymbol{r})\ , 
\end{align}\end{subequations} 
where it is in phase with $\boldsymbol{\dot{v}}$. For the phases in \eqref{klsjhgfdse} we will use the shorthand notation 
\begin{subequations}\label{ofdnjgkmf}\begin{align}
\begin{array}{ll} 
 &\text{exponent of }\widetilde{\boldsymbol{\dot{v}}},\ \widetilde{\boldsymbol{E}}_r,\ \widetilde{\boldsymbol{\dot{v}}}_k,\ \widetilde{\boldsymbol{E}}_a,\ \widetilde{\boldsymbol{\dot{v}}}\\ 
\widetilde{\boldsymbol{\dot{v}}}(\boldsymbol{r})&  -i\omega t_r\\ 
\widetilde{\boldsymbol{E}}_r(\boldsymbol{r}+\boldsymbol{n}R)& -i\omega (t_r+R/c)\\ 
\widetilde{\boldsymbol{\dot{v}}}_k(\boldsymbol{x}_k)& -i\omega (t_r+R/c)\\ 
\widetilde{\boldsymbol{E}}_a(\boldsymbol{x}_k-\boldsymbol{n}R)& -i\omega (t_r+R/c-R/c)\\  
\widetilde{\boldsymbol{\dot{v}}}(\boldsymbol{r})& -i\omega t_r\ .\end{array}
\end{align} 
If the fields do not propagate in vacuum, but in a medium with refractive index $n\neq 1$ (don't confuse this with the unit vector $\boldsymbol{n}$), then the phase relations are the following: 
\begin{align}
\begin{array}{ll} 
 &\text{exponent of }\widetilde{\boldsymbol{\dot{v}}},\ \widetilde{\boldsymbol{E}}_r,\ \widetilde{\boldsymbol{\dot{v}}}_k,\ \widetilde{\boldsymbol{E}}_a,\ \widetilde{\boldsymbol{\dot{v}}}\\ 
\widetilde{\boldsymbol{\dot{v}}}(\boldsymbol{r})&  -i\omega t_r\\ 
\widetilde{\boldsymbol{E}}_r(\boldsymbol{r}+\boldsymbol{n}R)& -i\omega (t_r+Rn/c)\\ 
\widetilde{\boldsymbol{\dot{v}}}_k(\boldsymbol{x}_k)& -i\omega (t_r+Rn/c)\\ 
\widetilde{\boldsymbol{E}}_a(\boldsymbol{x}_k-\boldsymbol{n}R)& -i\omega (t_r+Rn/c-Rn/c)\\  
\widetilde{\boldsymbol{\dot{v}}}(\boldsymbol{r})& -i\omega t_r \end{array}\label{ofdnjgkmfb}
\end{align} 
The advanced field $\widetilde{\boldsymbol{E}}_a$ is still in phase with the acceleration $\widetilde{\boldsymbol{\dot{v}}}$ at the position of the source $q$, and consequently is acting as an amplifying force, but not as a damping radiation\bz{-}back\bz{-}reaction. To achieve the necessary phase difference between $\widetilde{\boldsymbol{\dot{v}}}$ and $\widetilde{\boldsymbol{E}}_a$, Wheeler and Feynman decided for the following 
\begin{align}
&\text{\bfseries Postulate:}\left\{\mbox{\parbox{.72\linewidth}{$n\neq 1$ applies for retarded interactions.\newline{}$n=1$ applies for advanced interactions.}}\right.\notag\\ 
&\hspace{3em}\begin{array}{ll} 
 &\text{exponent of }\widetilde{\boldsymbol{\dot{v}}},\ \widetilde{\boldsymbol{E}}_r,\ \widetilde{\boldsymbol{\dot{v}}}_k,\ \widetilde{\boldsymbol{E}}_a,\ \widetilde{\boldsymbol{\dot{v}}}\\ 
\widetilde{\boldsymbol{\dot{v}}}(\boldsymbol{r})&  -i\omega t_r\\ 
\widetilde{\boldsymbol{E}}_r(\boldsymbol{r}+\boldsymbol{n}R)& -i\omega (t_r+Rn/c)\\ 
\widetilde{\boldsymbol{\dot{v}}}_k(\boldsymbol{x}_k)& -i\omega (t_r+Rn/c)\\ 
\widetilde{\boldsymbol{E}}_a(\boldsymbol{x}_k-\boldsymbol{n}R)& -i\omega (t_r+Rn/c-R/c)\\  
\widetilde{\boldsymbol{\dot{v}}}(\boldsymbol{r})& -i\omega\Big( t_r+R(n-1)/c\Big)\ .\end{array}\label{ofdnjgkmfc}
\end{align}\end{subequations} 
We will discuss this perplexing postulate in section~\ref{absch:why}\,. For the moment being we simply accept it, so that we can proceed with the presentation of absorber theory. Applying these phase relations, it is indeed possible to derive the desired expression for the radiation back\bz{-}reaction. For that purpose we first consider as absorber a plasma with a density of $N$ free electrons and $N$ ions per volume, which is surrounding the original source $q$ uniformly from all sides. The total advanced field radiated by all absorber particles at the position of the charge $q$ is 
\begin{align}
\boldsymbol{E}_a^{\text{total}}(t_r,\boldsymbol{r})&=\inte _0^\infty\!\dif R\inte _0^{2\pi}\!\dif\varphi\inte _0^\pi\! R^2\sin\vartheta \dif\vartheta\,\cdot N\boldsymbol{E}_a\ .   
\end{align} 
We define for this integration spherical coordinates with origin at the position $\boldsymbol{r}$ of the charge $q$. The polar axis is defined by $\boldsymbol{\dot{v}}$, as indicated in figure~\ref{fig:wheelfeyn2} \vpageref{fig:wheelfeyn2}. $\boldsymbol{E}_a=\eqref{oasaghdews}$ are the contributions of the single absorber particles (\ggie the electrons of the plasma), which we decompose into those components, which are parallel to $\boldsymbol{\dot{v}}$, and those components which are perpendicular to $\boldsymbol{\dot{v}}$: 
\begin{align}
\boldsymbol{E}_a&=\eqref{oasaghdews}=\boldsymbol{E}_a^{\parallel\boldsymbol{\dot{v}}} +\boldsymbol{E}_a^{\perp\boldsymbol{\dot{v}}}=\frac{\boldsymbol{\dot{v}}}{|\boldsymbol{\dot{v}}|}\,\frac{(\boldsymbol{E}_a\boldsymbol{\cdot\dot{v}})}{|\boldsymbol{\dot{v}}|}+\boldsymbol{E}_a^{\perp\boldsymbol{\dot{v}}}\stackrel{\text{figure }\ref{fig:wheelfeyn2}}{=}\notag\\ 
&=-\frac{\boldsymbol{\dot{v}}}{|\boldsymbol{\dot{v}}|}\, |\boldsymbol{E}_a|\,\sin\vartheta +\boldsymbol{E}_a^{\perp\boldsymbol{\dot{v}}} 
\stackrel{\eqref{oasaghdews}}{=}-\boldsymbol{\dot{v}}\,\frac{1}{(4\pi\epsilon _0)^2}\, \frac{e^2q\,\sin ^2\vartheta }{2m_eR^2c^4}+\boldsymbol{E}_a^{\perp\boldsymbol{\dot{v}}} \notag 
\end{align} 
$m_e$ is the electron mass, $-e$ is it's charge. $m_e$ and $-e$ now replace $m_k$ and $q_k$ of the previous formulas. Upon integration, the components perpendicular to $\boldsymbol{\dot{v}}$ mutually compensate. Therefore the integral can be written in the form 
\begin{align}
\boldsymbol{E}_a^{\text{total}}(t_r,\boldsymbol{r})&=-\boldsymbol{\dot{v}}\inte _0^\infty\!\dif R\inte _0^{2\pi}\!\dif\varphi\inte _0^\pi\! R^2\sin\vartheta \dif\vartheta\,\cdot N\,\cdot\notag\\ 
&\qquad\cdot\frac{1}{(4\pi\epsilon _0)^2}\, \frac{e^2q\,\sin ^2\vartheta}{2m_eR^2c^4}\ . 
\end{align} 
Obviously the integral over $R$ is diverging. How can that be? If the source is radiating only a finite amount of retarded energy, then the absorber's answer impossibly can be an infinite amount of advanced energy. We already indicated in figure~\ref{fig:phasen} the cause of this problem: We are assuming in our computations, that the amplitudes of the radiation fields are decreasing as $R^{-1}$ over arbitrary distances, even if appreciable amounts of energy have already been absorbed by particles which are nearer to the source. That's not correct. In an absorber, in which everywhere particles are extracting energy from the field, the field amplitude clearly will decrease faster than $\sim R^{-1}$. 

Wheeler and Feynman allow for that due to the insertion of a small coefficient of extinction 
\begin{align}
\kappa\in\mathbbm{R}\quad ,\quad 0<\kappa\ll 1\label{osjgfjhsd}
\end{align} 
into the computations. Clearly we must take care to insert $\kappa $ into advanced and retarded fields respectively with the appropriate signs, such that the result will be damping but not amplification of the fields. 

Could possibly the factor  
\begin{align}
\frac{\eqref{ofdnjgkmfc}}{\eqref{ofdnjgkmfb}}=\exp\{ -i\omega R(n-1)/c\} &\stackrel{\boldsymbol{\displaystyle ?}}{=}\exp\{ -\omega R\kappa /c\}\quad\text{is wrong\ARZp}\notag\\ 
\Longrightarrow\quad i(n-1)&\stackrel{\boldsymbol{\displaystyle ?}}{=}\kappa \quad\text{is wrong\ARZp} 
\end{align}
which Wheeler and Feynman got due to the replacement of the plausible assumption \eqref{ofdnjgkmfb} by the perplexing postulate \eqref{ofdnjgkmfc}, be somehow procured from the coefficient of extinction $\kappa $? This is not possible, however, because $i(n-1)$ is purely imaginary, while $\kappa $ must be real. 
 
The frequency\bz{-}dependent relative dielectric constant of a plasma is\!\cite[(7.59),(7.60)]{Jackson:electrodyn3}
\begin{align}
\epsilon _r(\omega )=1-\frac{\omega _{\text{plasma}}^2}{\omega ^2}\quad\text{if }\omega >\omega _{\text{plasma}}\equiv\sqrt{\frac{Ne^2}{\epsilon _0m_e}}\ . 
\end{align} 
Due to the $m_e\ll m_{\text{ion}}$ in the denominator of $\omega _{\text{Plasma}}$, the contribution of the ions to the dielectric properties of the plasma is negligible. The frequency\bz{-}dependent refractive index $n(\omega )$ of the plasma, which for simplicity we will assume to be non\bz{-}magnetic ($\mu _r=1$), therefore is 
\begin{align}
n(\omega )&=\sqrt{\epsilon _r\mu _r}=\sqrt{\epsilon _r} =\sqrt{1-\frac{Ne^2}{\epsilon _0m_e\omega ^2}}\notag\\ 
&\approx 1-\frac{Ne^2}{2\epsilon _0m_e\omega ^2}\quad\text{if }\omega\gg\omega _{\text{plasma}}\ .\label{kshgdsag}
\end{align} 
Adding the small extinction coefficient $\kappa $ to the refractive index, we get the complex frequency\bz{-}dependent refractive index 
\begin{align}
n_c=n-i\kappa =1-\frac{Ne^2}{2\epsilon _0m_e\omega ^2}-i\kappa\ .\label{skanhjgvkjh} 
\end{align} 
Now the integrals can be computed: 
\begin{align}
\widetilde{\boldsymbol{E}}_a^{\text{total}}(\omega ,\boldsymbol{r})&=-\widetilde{\boldsymbol{\dot{v}}}(\omega )\,\frac{e^2qN}{2m_ec^4(4\pi\epsilon _0)^2}\underbrace{\inte _0^{2\pi}\!\dif\varphi\inte _0^\pi\!\dif\vartheta\,\sin ^3\vartheta}_{\displaystyle 8\pi /3}\,\cdot\notag\\ 
&\qquad\cdot\underbrace{\inte _0^\infty\!\dif R\,\exp\{ -i\omega R(n-i\kappa -1)/c\}}_{\displaystyle\hspace{-4em}\Big[ \frac{2\epsilon _0m_e\omega c}{iNe^2}\,\exp\{ -i\omega R(n-i\kappa -1)/c\}\Big] _0^\infty\hspace{-4em}}\notag\\ 
&=\frac{\omega\widetilde{\boldsymbol{\dot{v}}}}{i}\,\frac{2q }{3c^3 4\pi\epsilon _0}
\stackrel{\eqref{lsdghjsavc}}{=}\frac{2q }{3c^3 4\pi\epsilon _0}\,\widetilde{\boldsymbol{\ddot{v}}}(\omega )\notag\displaybreak[1]\\ 
\boldsymbol{E}_a^{\text{total}}(t_r,\boldsymbol{r})&=\frac{2q }{3c^3 4\pi\epsilon _0}\,\boldsymbol{\ddot{v}}(t_r)\label{kdjhsddgsg}
\end{align} 
The very small terms $\sim\!\kappa $ could be skipped in the linear part of the equation (but not in the exponent\ARZp ). Compare this result to the radiation back\bz{-}reaction indicated by Abraham and Lorentz: 
\begin{align}
\boldsymbol{F}_{\text{rad}}\stackrel{\eqref{ksdgvsgc}}{=}\frac{2q^2\boldsymbol{\ddot{v}}}{3c^34\pi\epsilon _0}\quad\text{if }v\ll c 
\end{align} 
Wheeler and Feynman comment: \al We conclude that the force of radiative reaction arises, not from the direct action of a particle upon itself, but from the advanced action upon this charge caused by the future motion of the particles of the absorber.\ar 

Wheeler and Feynman make sure, that the result is the same, if the absorber is a material with bound electrons. The refractive index of such materials is $n>1$. To get a model for this case, they insert into equation \eqref{oiskgjhs} a complex function $p$, which does depend not only on the absorber's material, but also on the radiation's frequency $\omega $: 
\begin{subequations}\label{kshgskggs}\begin{align}
\boldsymbol{\dot{v}}_k\stackrel{\eqref{oiskgjhs}}{=}\frac{q_k\boldsymbol{E}_r(x)}{m_k}\quad\longrightarrow\quad 
\boldsymbol{\dot{v}}_k=\frac{q_k\boldsymbol{E}_r(x)}{m_k}\cdot p(\omega ) 
\end{align} 
In case of very high frequencies $\omega $ and\bz{/}or very weakly bound electrons in the absorber, $p$ will approach 1, and we get the limit of the plasma considered before. In other cases, $p$ will differ from 1, and may even have a significant imaginary part: 
\begin{align}
n_c(\omega )=\eqref{skanhjgvkjh}\quad \longrightarrow\quad  
&n(\omega )-i\kappa =\sqrt{1-\frac{Ne^2p(\omega )}{\epsilon _0m_e\omega ^2}}\label{ppfbjndljbhds}\\ 
&n,\kappa\in\mathbbm{R}\notag 
\end{align}\end{subequations} 
\begin{figure}[!b]\centering\begin{overpic}{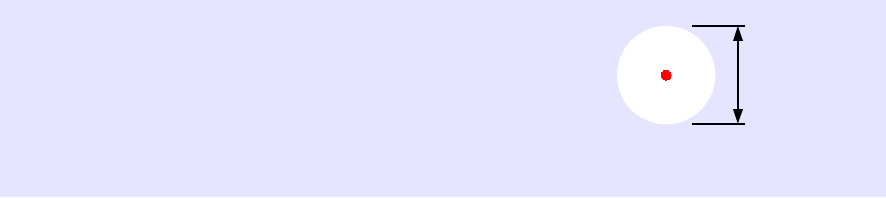}
\put(85,13){$2R_{\text{cav}}$}
\put(77,13){$q$}
\put(10,15){\textcolor{blue}{$N$ absorber particles\,/\,volume}}
\end{overpic}\caption{Cavity in the continuous absorber medium}\label{fig:kavitaet}\end{figure} 
To arrive \emph{exactly} at the desired result for radiation back\bz{-}reaction, Wheeler and Feynman applied a little trick: They did not embed the source $q$ directly into the absorber medium, but put it --- as indicated in figure~\ref{fig:kavitaet} --- into the center of a small spherical cavity with radius $R_{\text{cav}}$. The effect of that trick is this: Within the cavity, the index of refraction is 1 (vacuum), while in the absorber medium the refractive index is $n\neq 1$. Consequently there will be a reflection at the cavity's surface: 
\begin{align}
\Bigg|\frac{\widetilde{\boldsymbol{E}}_a^{\text{total, reflected}}}{\widetilde{\boldsymbol{E}}_a^{\text{total, incoming}}}\Bigg| &=\frac{|1-n+i\kappa |}{1+n-i\kappa }\notag\displaybreak[1]\\ 
\Longrightarrow\quad\frac{\widetilde{\boldsymbol{E}}_a^{\text{total, transmitted}}}{\widetilde{\boldsymbol{E}}_a^{\text{total, incoming}}}&=\frac{2}{1+n-i\kappa }\quad\text{if }n\geq 1\label{ksksdhndgg}
\end{align} 
The factor $\eqref{ksksdhndgg}\approx 1$ will bring about --- as we will see immediately --- a result which is not only excellent, but perfect.\enlargethispage{1\baselineskip} 

Only the retarded field emitted by $q$ is subject to reflection, as Wheeler and Feynman again assume \eqref{ofdnjgkmfc}: The retarded field propagates from $q$ to the absorber particles in the medium with a refractive index $n_c=\eqref{ppfbjndljbhds}$, while the advanced fields are propagating from the absorber particles to the source $q$ in vacuum with $n=1$. Inserting \eqref{kshgskggs}, we consequently get: 
\vspace{-4ex}\begin{align} 
&\widetilde{\boldsymbol{E}}_a^{\text{total}}(\omega ,\boldsymbol{r})\stackrel{\eqref{kdjhsddgsg}}{=}-\widetilde{\boldsymbol{\dot{v}}}(\omega )\,\frac{pe^2qN}{2m_ec^4(4\pi\epsilon _0)^2}\overbrace{\inte _0^{2\pi}\!\dif\varphi\inte _0^\pi\!\dif\vartheta\,\sin ^3\vartheta}^{\displaystyle 8\pi /3}\,\cdot\notag\\ 
&\cdot\frac{2}{1+n-i\kappa}\,\underbrace{\inte _{R_{\text{cav}}}^{\infty}\!\dif R\,\exp\Big\{ -i\,\frac{(R-R_{\text{cav}})\,\omega}{c}\, (n-i\kappa -1)\Big\}}_{\displaystyle\hspace{-8em}\Big[ -\frac{c}{i\omega\, (n-i\kappa -1)}\,\exp\Big\{ -i\,\frac{(R-R_{\text{cav}})\,\omega}{c}\, (n-i\kappa -1)\Big\}\Big] _{R_{\text{cav}}}^\infty\hspace{-6em}}\label{posahnbgkwr} 
\end{align} 
Observing 
\begin{align}
(1+n-i\kappa )(-1+n-i\kappa )=-1+(n-i\kappa )^2\stackrel{\eqref{ppfbjndljbhds}}{=}-\frac{Ne^2p}{\epsilon _0m_e\omega ^2}\ ,\notag  
\end{align} 
the result becomes 
\begin{align} 
\widetilde{\boldsymbol{E}}_a^{\text{total}}(\omega ,\boldsymbol{r})&=+i\omega\widetilde{\boldsymbol{ \dot{v}}}\,
\underbrace{\frac{Ne^2p}{\epsilon _0m_e\omega ^2(1+n-i\kappa )(n-i\kappa -1)}}_{-1}\,\frac{2q}{3c^3 4\pi \epsilon _0} \notag\\ 
\ggstackrel[.2]{\eqref{lsdghjsavc}}+\widetilde{\boldsymbol{ \ddot{v}}}\,\frac{2q}{3c^3 4\pi \epsilon _0}\notag\\ 
\boldsymbol{E}_a^{\text{total}}(t_r,\boldsymbol{r})&=\frac{2q}{3c^3 4\pi\epsilon _0}\,\boldsymbol{\ddot{v}}(t_r)\ .\label{sjgksdghsfg}
\end{align} 
The result is independent of the material of the absorber, because the function $p(\omega )$ cancels. For a medium with arbitrary frequency\bz{-}dependent $p(\omega )\neq 1$ we get the same result as for the plasma. Remember, however, that this beautiful result again is based on the strange postulate \eqref{ofdnjgkmfc}, which will be discussed in section~\ref{absch:why}. 

\begin{figure}[!htb]\centering\begin{overpic}{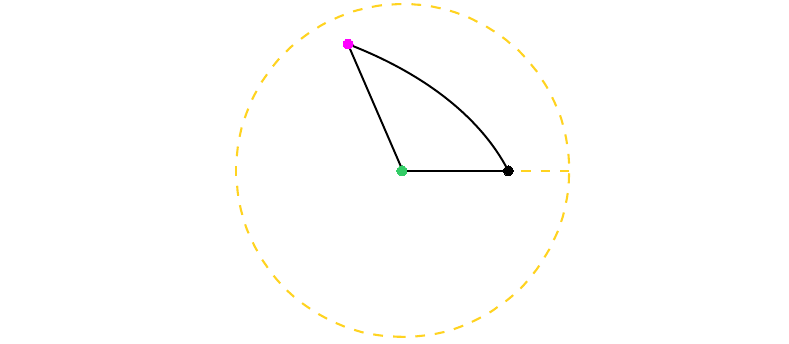}
\put(39.5,36){\textcolor{magenta}{$\boldsymbol{n}$}}
\put(44,27){$\vartheta $}
\put(52,24){$\varphi $}
\put(56,18){$\sigma $}
\put(60,17){$\boldsymbol{s}/|\boldsymbol{s}|$}
\put(52.5,36.5){\rotatebox{-37}{$(\boldsymbol{s},\boldsymbol{n})$}}
\put(41,17){\textcolor{dklgrn}{$\boldsymbol{\dot{v}}/|\boldsymbol{\dot{v}}|$}}
\end{overpic}\caption{The system of spherical coordinates}\label{fig:sphcoor}\end{figure}
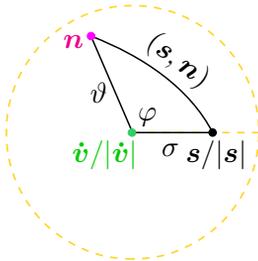

Now we are going to investigate the value of $\widetilde{\boldsymbol{E}}_a^{\text{total}}$ at a position $\boldsymbol{r}+\boldsymbol{s}$ in the cavity of figure~\ref{fig:kavitaet}, which is much closer to the position $\boldsymbol{r}$ of the source than the nearest absorber particle, such that 
\begin{align}
s\equiv|\boldsymbol{s}|\ll R_{\text{cav}}
\end{align} 
holds. The field $\widetilde{\boldsymbol{E}}_a(\omega )$ emitted by a certain absorber particle, which is located in direction $\boldsymbol{n}$ (as seen from the position of the source), differs at position $\boldsymbol{r}+\boldsymbol{s}$ from the field at the position $\boldsymbol{r}$ of the source $q$ by the phase\bz{-}factor 
\begin{align}
\exp\Big\{ -i\omega\,\frac{s\cos (\boldsymbol{s},\boldsymbol{n})}{c}\Big\}\ . \label{ksgdhsddk} 
\end{align} 
(Remember that the points $\boldsymbol{r}+\boldsymbol{s}$ and $\boldsymbol{r}$ both are located in the cavity of the absorber, where the refractive index is $n=1$.) Because of $s\ll R_{\text{cav}}$, in comparison to \eqref{posahnbgkwr} only the integrals over $\varphi $ and $\vartheta $ are different, but not the integral over $R$: 
\begin{subequations}\begin{align}
&\widetilde{\boldsymbol{E}}_a^{\text{total}}(\omega ,\boldsymbol{r}+\boldsymbol{s})=\widetilde{\boldsymbol{\ddot{v}}}(\omega ,\boldsymbol{r})\,\frac{2q}{3c^34\pi\epsilon _0}\cdot I(\omega ,\boldsymbol{r}+\boldsymbol{s}) \\ 
&I\equiv\frac{3}{8\pi}\inte _0^{2\pi}\!\dif\varphi\inte _0^\pi\!\dif\vartheta\,\sin ^3\vartheta\,\exp\Big\{ -i\omega\,\frac{s\cos (\boldsymbol{s},\boldsymbol{n})}{c}\Big\} \label{klshgsdgvsdb}   
\end{align}\end{subequations} 
We continue to use spherical coordinates with the origin located at the position of the charge $q$. The polar axis is oriented parallel to the acceleration $\boldsymbol{\dot{v}}(t_r,\boldsymbol{r})$ of this charge. Until now we did not fix the zero\bz{-}point of the azimuthal angle $\varphi $. We do this now by the stipulation, that the point $\boldsymbol{r}+\boldsymbol{s}$ shall have the spherical coordinates $(\, |\boldsymbol{s}|,\vartheta =\sigma ,\varphi =0\, )$, see \mbox{fig.\,\ref{fig:sphcoor}} \vpageref[\negmedspace ]{fig:sphcoor}. 
In this figure we are looking against the direction of the polar axis onto a sphere with radius 1. The center of the sphere is identical with the origin of the spherical coordinates. The dashed orange circle is the sphere's equator ($\vartheta =\pi /2$), the dashed straight orange line (a large part of which is covered by the black line $\sigma $) is the zero\bz{-}meridian ($\varphi =0$). The point $\boldsymbol{\dot{v}}/|\boldsymbol{\dot{v}}|$ is located at the sphere's north pole ($\vartheta =0$). The coordinates of point $\boldsymbol{n}$ are $(1,\vartheta ,\varphi )$, and $\vartheta $ is the length of the meridian\bz{-}section, which is the shortes connection (geodesic) on the sphere's surface inbetween $\boldsymbol{\dot{v}}/|\boldsymbol{\dot{v}}|$ and $\boldsymbol{n}$. The length of the geodesic inbetween $\boldsymbol{\dot{v}}/|\boldsymbol{\dot{v}}|$ and the point $\boldsymbol{s}/|\boldsymbol{s}|$ is $\sigma $. During the integration \eqref{klshgsdgvsdb}, the point $\boldsymbol{n}$ is running over all the sphere's surface. The angle $(\boldsymbol{s},\boldsymbol{n})$, which is identical to one side of the triangle sketched in \mbox{fig.\,\ref{fig:sphcoor}}, can be computed by means of the cosine theorem of spherical trigonometry (see \ggeg\!\!\cite[part\,II,IV.B.7.]{Bronstein:Taschenbuch}): 
\begin{align}
\cos (\boldsymbol{s},\boldsymbol{n})=\cos\vartheta\,\cos\sigma +\sin\vartheta\,\sin\sigma\,\cos\varphi 
\end{align}
This is inserted into the integral \eqref{klshgsdgvsdb}: 
\begin{align}
I&\equiv\frac{3}{4}\inte _0^\pi\!\dif\vartheta\,\sin ^3\vartheta\,\exp\Big\{ -i\omega\,\frac{s\cos\vartheta\,\cos\sigma}{c}\Big\}\,\cdot\notag\\ 
&\qquad\cdot\frac{1}{2\pi}\inte _0^{2\pi}\!\dif\varphi\,\exp\Big\{ -i\omega\,\frac{s\sin\vartheta\,\sin\sigma\,\cos\varphi}{c}\Big\}  
\end{align} 
The integral over $\varphi $ is the Bessel function 
\begin{align}
&\frac{1}{2\pi}\inte _0^{2\pi}\!\dif\varphi\,\exp\Big\{ -i\omega\,\frac{s\sin\vartheta\,\sin\sigma\,\cos\varphi}{c}\Big\}\, =\notag\\ 
&=J_0(-i\omega\, s\sin\vartheta\,\sin\sigma /c)= 
\sum _{j=0}^\infty \frac{(-1)^j(-i\omega\, s\sin\vartheta\,\sin\sigma /c)^{2j}}{j!\,2^{2j}\Gamma (j+1)}\ .\notag   
\end{align} 
The integral over $\vartheta $, whose integrand is the product of the exponential function and the Bessel function, which again due to the trigonometric functions depends on $\vartheta $, leads to a quite complicated combination of hypergeometric series. There exists, however, a much simpler method of solution, which is based on an appropriate change of coordinates. Even though Wheeler and Feynman apply that transformation of coordinates, they still present in the end for mysterious reasons the result in form of hypergeometric series\!\cite[(14) to (17)]{Wheeler:absorber}. Probably they tried --- like me --- both methods of solution, and eventually in error printed in their publication a mix of both methods. 

\begin{figure}[!t]\centering\begin{overpic}{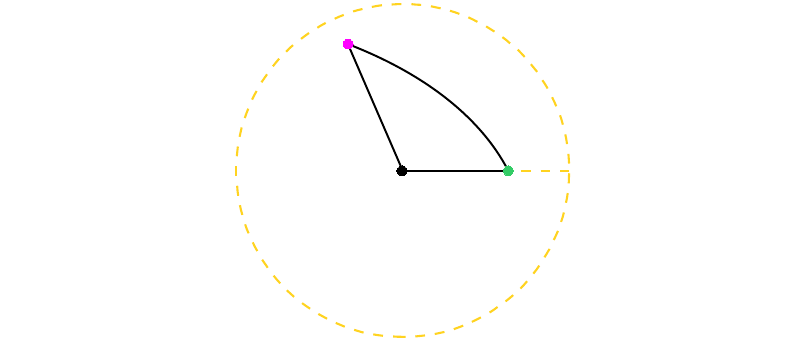}
\put(39.5,36){\textcolor{magenta}{$\boldsymbol{n}$}}
\put(44,27){$\vartheta $}
\put(52,24){$\varphi $}
\put(56,18){$\sigma $}
\put(60,17){\textcolor{dklgrn}{$\boldsymbol{\dot{v}}/|\boldsymbol{\dot{v}}|$}}
\put(52.5,36.5){\rotatebox{-37}{$(\boldsymbol{\dot{v}},\boldsymbol{n})$}}
\put(41,17){$\boldsymbol{s}/|\boldsymbol{s}|$}
\end{overpic}\caption{The system of spherical coordinates}\label{fig:sphcoor2}\end{figure}

This is the transformation of coordinates, which paves the way to the simple solution: The origin of the spherical coordinates stays at the position $\boldsymbol{r}$ of the charge $q$. The polar axis, which until now was oriented parallel to the acceleration $\boldsymbol{\dot{v}}(t_r,\boldsymbol{r})$ of this charge, now will be oriented parallel to the vector $\boldsymbol{s}$. Thereby the angle $(\boldsymbol{s},\boldsymbol{n})$ of \eqref{ksgdhsddk} now becomes $\vartheta $, and on the other hand $\sin ^2\vartheta $ now becomes $\sin ^2(\boldsymbol{\dot{v}},\boldsymbol{n})$:  
\begin{subequations}\begin{align}
&\widetilde{\boldsymbol{E}}_a^{\text{total}}(\omega ,\boldsymbol{r}+\boldsymbol{s})=\widetilde{\boldsymbol{\ddot{v}}}(\omega ,\boldsymbol{r})\,\frac{2q}{3c^34\pi\epsilon _0}\cdot I(\omega ,\boldsymbol{r}+\boldsymbol{s})\label{klshgsdgea} \\ 
&I\equiv\frac{3}{8\pi}\inte _0^{2\pi}\!\dif\varphi\inte _0^\pi\!\dif\vartheta\,\sin\vartheta\,\sin ^2(\boldsymbol{\dot{v}},\boldsymbol{n})\,\exp\Big\{ -i\omega\,\frac{s\cos\vartheta}{c}\Big\} \label{klshgsdgeb}   
\end{align}\end{subequations}
The zero\bz{-}point of the azimuthal angle $\varphi $ is fixed by the stipulation, that the sperical coordinates of $\boldsymbol{\dot{v}}$ shall be $(\, |\boldsymbol{\dot{v}}|\, ,\,\vartheta =\sigma\, ,\,\varphi =0\, )$, see \mbox{fig.\,\ref{fig:sphcoor2}}. 
Now the point $\boldsymbol{s}/|\boldsymbol{s}|$ is located at the north pole ($\vartheta =0$) of the sphere with radius 1. The coordinates of the point $\boldsymbol{n}$ are $(1,\vartheta ,\varphi )$, as before. $\vartheta $ now is the length of the meridian\bz{-}section between $\boldsymbol{s}/|\boldsymbol{s}|$ and $\boldsymbol{n}$. The length of the geodesic inbetween $\boldsymbol{s}/|\boldsymbol{s}|$ and the point $\boldsymbol{\dot{v}}/|\boldsymbol{\dot{v}}|$ is $\sigma $. The angle $(\boldsymbol{\dot{v}},\boldsymbol{n})$, which is identical to one side of the triangle sketched in \mbox{fig.\,\ref{fig:sphcoor2}} can again be determined by means of the cosine theorem of spherical trigonometry: 
\begin{align}
\cos (\boldsymbol{\dot{v}},\boldsymbol{n})=\cos\vartheta\,\cos\sigma +\sin\vartheta\,\sin\sigma\,\cos\varphi 
\end{align}
Using the relation $\sin ^2\vartheta =1-\cos ^2\vartheta $, this is inserted into the integral \eqref{klshgsdgeb}: 
\begin{align}
I&=\frac{3}{8\pi}\inte _0^\pi\!\dif\vartheta\,\sin\vartheta\,\exp\Big\{ -i\omega\,\frac{s\cos\vartheta}{c}\Big\}\cdot\inte _0^{2\pi}\!\dif\varphi\,\Big( 1-\cos ^2\vartheta\,\cos ^2\sigma\, -\notag\\ 
&\qquad -2\cos\vartheta\,\cos\sigma\,\sin\vartheta\,\sin\sigma\,\cos\varphi -\sin ^2\vartheta\,\sin ^2\sigma\,\cos ^2\varphi \Big)\notag\\  
&=\frac{3}{8}\inte _0^\pi\!\dif\vartheta\,\sin\vartheta\,\exp\Big\{ -i\omega\,\frac{s\cos\vartheta}{c}\Big\}\,\cdot\notag\\ 
&\qquad\cdot\Big( 2-2\cos ^2\vartheta\,\cos ^2\sigma -\sin ^2\vartheta\,\sin ^2\sigma \Big) 
\end{align} 
Applying the substitution 
\begin{align}
x\equiv\cos\vartheta\qquad ,\qquad \inte _{+1}^{-1}\!\dif x=-\inte _{0}^{\pi}\!\dif\vartheta\,\sin\vartheta\ ,  
\end{align}
$I$ is transformed into 
\begin{subequations}\begin{align}
I&=-\frac{3}{4}\inte _{+1}^{-1}\!\dif x\,\exp\Big\{ -iux\Big\}\,\cdot\Big( 1-y -x^2(1-3y)\,\Big) \\ 
u&\equiv\frac{\omega s}{c}\qquad ,\qquad y\equiv\frac{\sin ^2\sigma}{2}\ ,   
\end{align}\end{subequations} 
and can now be solved easily: 
\begin{align}
I&=-\frac{3(1-y)}{4}\,\frac{i}{u}\,\exp\{ -iux\}\Big| _{+1}^{-1}\, +\notag\\ 
&\quad +\frac{3(1-3y)}{4}\,\Big( \frac{ix^2}{u}+\frac{2x}{u^2}-\frac{i2}{u^3}\Big)\,\exp\{ -iux\} \bigg| _{+1}^{-1} \notag\\ 
&=\frac{3}{4}\,\Big[ -2y\,\frac{i}{u}+(1-3y)\,\Big( -\frac{2}{u^2}-\frac{i2}{u^3}\Big)\,\Big]\,\exp\{ +iu\} \notag\\ 
&\quad -\frac{3}{4}\,\Big[ -2y\,\frac{i}{u}+(1-3y)\,\Big( \frac{2}{u^2}-\frac{i2}{u^3}\Big)\,\Big]\,\exp\{ -iu\}\notag\\ 
&=\Big(\frac{3y}{u}+\frac{3(1-3y)}{u^3}\Big)\,\sin\{ +u\} -\frac{3(1-3y)}{u^2}\,\cos\{ +u\} \label{ksksghdgfs}
\end{align} 
The limit $I\rightarrow 1$ must hold in case $u\rightarrow 0$ \ggresp $s\rightarrow 0$, as can be read immediately from \eqref{klshgsdgvsdb}. By means of the rule of l'Hospital we doublecheck, whether our result is consistent in this respect: 
\begin{align}
\lim _{u\rightarrow 0}I&=\lim _{u\rightarrow 0}\frac{(3yu^2+3-9y)\sin u\, -(3-9y)u\cos u}{u^3}\notag\\ 
&=\lim _{u\rightarrow 0}\frac{yu^2\cos u\, +(1-y)u\sin u}{u^2}\notag\\ 
&=\lim _{u\rightarrow 0}\frac{(-yu^2+1-y)\sin u\, +(1+y)u\cos u}{2u}\notag\\ 
&=\lim _{u\rightarrow 0}\frac{(-2yu-1-y)u\sin u\, +(-yu^2+2)\cos u}{2}\notag\\ 
&=1\label{pjgfnjhsgdfsbg}
\end{align} 

The dimension\bz{-}less number $u$ is equal to the distance $s$ between the point of observation and the source $q$, divided by $(\text{wavelength}/2\pi )$. Therefore the term $\sim u^{-1}$ will dominate the result, provided that the distance $s$ is at least some few wavelengths. In this case 
\begin{align}
I&=\frac{3\sin ^2\sigma}{2} \,\frac{c}{\omega s}\,\frac{1}{2i}\,\Big[\,\exp\{ +i\omega s/c\} -\exp\{ -i\omega s/c\}\Big] \label{osajnhgds}\\ 
&= \frac{3\sin ^2\sigma}{2} \,\cdot\,\frac{\sin (\omega s/c)}{\omega s/c} \notag 
\end{align}
holds. We insert \eqref{osajnhgds} into \eqref{klshgsdgea}, and make use of 
\begin{align}
\widetilde{\boldsymbol{\ddot{v}}}(\omega )\stackrel{\eqref{lsdghjsavc}}{=}-i\omega\widetilde{\boldsymbol{\dot{v}}}\ ,\notag  
\end{align} 
to describe the field as a function of $\widetilde{\boldsymbol{\dot{v}}}$ (but not of $\widetilde{\boldsymbol{\ddot{v}}}$): 
\begin{subequations}\begin{align}
&\boldsymbol{E}_a^{\text{total}}(t,\boldsymbol{r}+\boldsymbol{s})=\inte _{-\infty}^{+\infty}\!\frac{\dif\omega}{2\pi}\,\widetilde{\boldsymbol{E}}_a^{\text{total}}(\omega ,\boldsymbol{r}+\boldsymbol{s})\,\exp\{ -i\omega t\}\, =\notag\\  
&=-\frac{\sin ^2\sigma}{2}\inte _{-\infty}^{+\infty}\!\frac{\dif\omega}{2\pi}\,\widetilde{\boldsymbol{\dot{v}}}(\omega ,\boldsymbol{r})\,\frac{q}{sc^24\pi\epsilon _0}\,\cdot\notag\\ 
&\qquad\cdot \Big[\,\exp\Big\{ -i\omega\,\Big( t-\frac{s}{c}\Big)\Big\} -\exp\Big\{ -i\omega\,\Big( t+\frac{s}{c}\Big)\Big\} \Big]\, =\notag\\ 
&=-\frac{\sin ^2\sigma}{2}\,\frac{q}{sc^24\pi\epsilon _0}\cdot\Big(\boldsymbol{\dot{v}}(t_r)-\boldsymbol{\dot{v}}(t_a)\Big)\label{klsnjhgdka}\\ 
&\qquad{\text{with }}t_r=t-s/c\ ,\ t_a=t+s/c\notag   
\end{align}
We want to compare this result with the retarded and advanced fields, which the accelerated source $q$ at position $\boldsymbol{r}$ is radiating to the position $\boldsymbol{r}+\boldsymbol{s}$. The radiated fields are in non\bz{-}relativistic approximation (the index $_s$ is coding for $_r=\text{retarded}$ or $_a=\text{advanced}$, and the upper one of double signs applies for the retarded case, the lower one for the advanced case) 
\begin{align}
\boldsymbol{E}_s(t,\boldsymbol{r}+\boldsymbol{s})\stackrel{\eqref{oshgkdsdgsda}}{=}\frac{1}{4\pi\epsilon _0}\,\frac{1}{s^2}\, \frac{q\boldsymbol{s\times}(\boldsymbol{s\times\dot{v}})}{sc^2}\Big| _{t_s=t\mp s/c}\ .\label{klsnjhgdkb}  
\end{align}
Here the unit vector $\boldsymbol{n}$ has been written in the form $\boldsymbol{s}/s$, and $s$ instead of $R$ was inserted for the distance between the source and the point of observation. The projection of these fields onto the axis of $\boldsymbol{\dot{v}}$ is 
\begin{align}
\boldsymbol{E}_{s\,\parallel}(t,\boldsymbol{r}+\boldsymbol{s}) &\equiv\frac{\boldsymbol{\dot{v}}\cos\Big(\boldsymbol{\dot{v}},\boldsymbol{s\times}(\boldsymbol{s\times\dot{v}})\Big)}{\dot{v}}\Big| _{t_s=t\mp s/c}\cdot\Big| \boldsymbol{E}_s(t,\boldsymbol{r}+\boldsymbol{s})\Big|\notag\\ 
&=\frac{\boldsymbol{\dot{v}}\cos\Big(\boldsymbol{\dot{v}},\boldsymbol{s\times}(\boldsymbol{s\times\dot{v}})\Big)}{\dot{v}}\,\frac{1}{4\pi\epsilon _0}\,\frac{q\dot{v}\sin (\boldsymbol{s},\boldsymbol{\dot{v}})}{sc^2}\Big| _{t_s=t\mp s/c} \notag 
\end{align} 
$(\boldsymbol{s},\boldsymbol{\dot{v}})=\sigma $ holds in our spherical coordinates (see \mbox{fig.\,\ref{fig:sphcoor2})}, and  
\begin{align}
\boldsymbol{s\times}(\boldsymbol{s\times\dot{v}})&=(\, s\dot{v}\, ,\,\vartheta =\pi /2\, ,\,\varphi =\pi\,)\notag\\ 
\cos\Big(\boldsymbol{\dot{v}},\boldsymbol{s\times}(\boldsymbol{s\times\dot{v}})\Big) &=\cos (\sigma +\pi /2)=-\sin\sigma
\ .\notag 
\end{align}
Thereby one gets 
\begin{align}
\boldsymbol{E}_{s\,\parallel}(t,\boldsymbol{r}+\boldsymbol{s}) &=-\sin ^2\sigma\,\frac{1}{4\pi\epsilon _0}\,\frac{q\boldsymbol{\dot{v}}}{sc^2}\Big| _{t_s=t\mp s/c}\ . \label{klsnjhgdkc}
\end{align} 
Thus the total advanced field \eqref{klsnjhgdka} radiated by all absorber particles from distances $R\geq R_{\text{cav}}\gg s$ is nearby the source $q$ identical to the difference between the projections onto the $\boldsymbol{\dot{v}}$\bz{-}axis of half the retarded field radiated by $q$ and half the advanced field radiated by $q$. Wheeler and Feynman comment: 
\begin{flushright}\parbox{.8\linewidth}{\al In words, formula (19) [that is our equation \eqref{klsnjhgdka}] states that the advanced field of the absorber is equal in the neighborhood of the accelerated particle to [\dots ] the difference between half the retarded field (first term) and half the advanced field (second term) which one calculates for the source itself.\ar}\parbox{.14\linewidth}{\begin{align}\label{lksakdsgjc}
\end{align}}\end{flushright}\end{subequations} 
This conclusion obviously is not correct, as the fields \eqref{klsnjhgdka} are oriented parallel (or antiparallel) to $\boldsymbol{\dot{v}}$, while the fields \eqref{klsnjhgdkb} have components oriented perpendicular to $\boldsymbol{\dot{v}}$, which are observable and different from zero. This discrepancy will be discussed in section~\ref{absch:error19}. 

\begin{figure}[!b]\centering\includegraphics{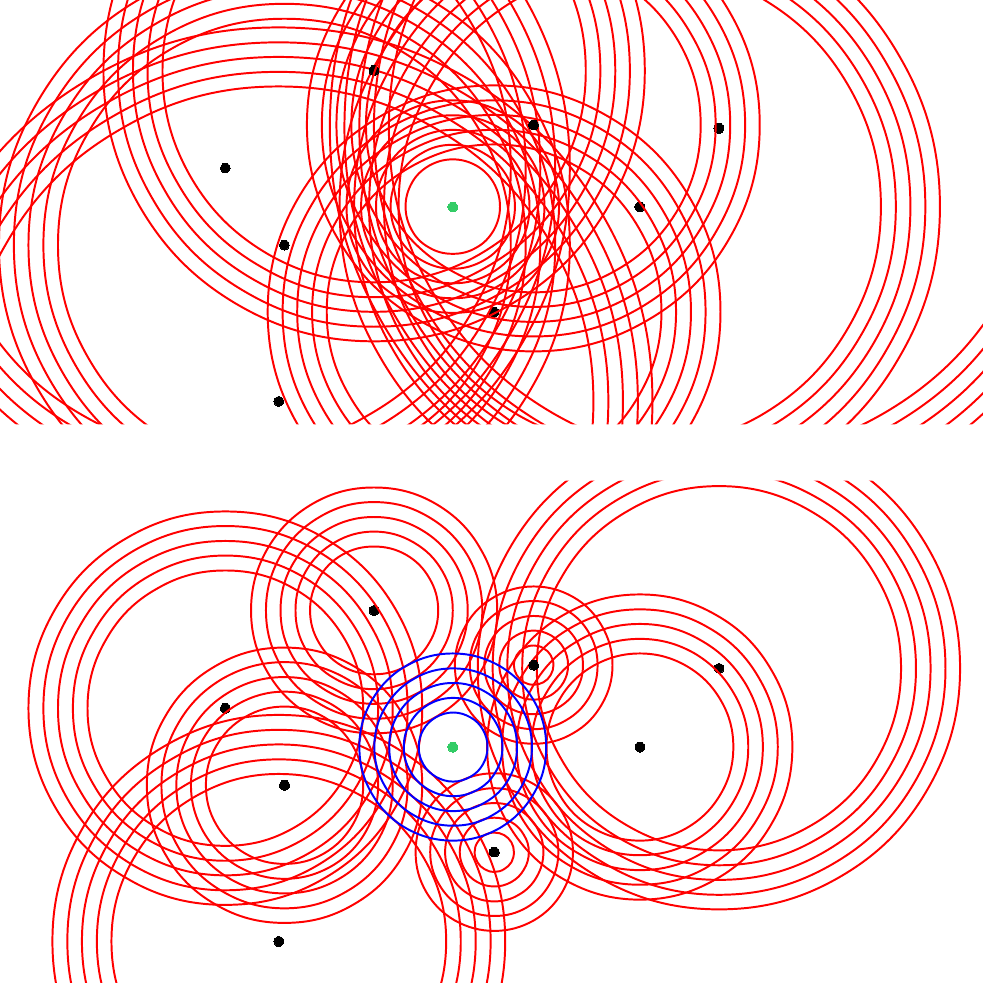}
\caption{\textcolor{red}{Advanced} and \textcolor{blue}{retarded} waves}\label{fig:wellen}\end{figure}

For the moment being we accept the assertion \eqref{lksakdsgjc}, and illustrate it in figure~\ref{fig:wellen} \vpageref[\negmedspace ]{fig:wellen}.
The upper part of the sketch is a snapshot shortly before (in sense of the usual time direction) the moment $t_r$, in which the source $q$ (indicated in the sketch by a green dot) will radiate a retarded field. Nine advanced wave\bz{-}trains can bee seen. Eight of them are collapsing onto one of eight absorber particles (indicated as black dots) each, and one --- namely the \al proper\ar  advanced field  \eqref{klsnjhgdkb} of the source $q$, whose phase is shifted by $\pi $ versus the other fields  --- is collapsing onto the primary source $q$. Viewed from nearby $q$, the superposition of the nine fields looks like one single advanced spherical wave, which is collapsing onto $q$. Wheeler and Feynman describe this scenario in their equation (24): 
\vspace{-2ex}\begin{align}
\left(\mbox{\parbox{22.5mm}{\raggedright\scriptsize\mbox{total disturbance} \mbox{converging on} \mbox{\,source}}}\right)\!  
=\!\left(\mbox{\parbox{21.5mm}{\raggedright\scriptsize\mbox{proper advanced} \mbox{field of source} \mbox{\,itself}}}\right)\!\!  
+\!\!\left(\mbox{\parbox{40mm}{\raggedright\scriptsize\mbox{field apparently convergent on} \mbox{source, actually composed of} \mbox{parts convergent on individual} \mbox{\,absorber particles}}}\right) \notag 
\end{align}\vspace{-2ex}\par\noindent 
The signs of the sum of the advanced field \eqref{klsnjhgdka} and the advanced field \eqref{klsnjhgdkb} emitted by $q$ are different, \ggie they are out of phase by $\pi $, and interfering destructively. As no experimenter ever observed an advanced field, the amplitudes of \eqref{klsnjhgdka} and \eqref{klsnjhgdkb} must be equal, such that they will mutually delete completely. 

The lower sketch is a snapshot shortly after $t_r$. The eight advanced waves of the absorber particles have run over the source $q$, and continue to collapse onto one absorber particle each. Their superposition is looking like a retarded wave emanated from $q$ and propagating into the future. At the same time, according to \eqref{klsnjhgdkb} a retarded wave emitted by $q$, which is displayed in blue color in fig.~\ref{fig:wellen}, is propagating into the future. This field is in phase and constructively interfering with the field \eqref{klsnjhgdka}, which is sketched in red. In their equation (20), Wheeler and Feynman describe this scenario by 
\vspace{-2ex}\begin{align}
\left(\mbox{\parbox{22.5mm}{\raggedright\scriptsize\mbox{total disturbance} \mbox{diverging from} \mbox{\,source}}}\right)\!  
=\!\left(\mbox{\parbox{21mm}{\raggedright\scriptsize\mbox{proper retarded} \mbox{field of source} \mbox{\,itself}}}\right)\!\!  
+\!\!\left(\mbox{\parbox{41mm}{\raggedright\scriptsize\mbox{field apparently diverging from} \mbox{source, actually composed of} \mbox{parts converging on individual} \mbox{\,absorber particles}}}\right).\notag 
\end{align}
The signs of the advanced fields \eqref{klsnjhgdka} and of the retarded field \eqref{klsnjhgdkb} emitted by the source $q$ are equal, \ggie they are in phase and interfering constructively. Only the sum of all the fields can be measured; the experimenter cannot discern the contributions to the total field coming from the fields \eqref{klsnjhgdka} and from the field \eqref{klsnjhgdkb} respectively. We know from \eqref{klsnjhgdkb}, however, that the amplitudes of the retarded and the advanced fields emitted by $q$ are equal. And we just stated that the amplitude of the advanced field emitted by $q$ must be equal to the amplitude of \eqref{klsnjhgdka}. From this we conclude, that in both the spherical wave (seemingly) collapsing upon $q$ and the spherical wave (seemingly) emanating form $q$, half of the amplitude must be contributed by the proper field \eqref{klsnjhgdkb} of the source, and the other half of the amplitude must be contributed by the superposition \eqref{klsnjhgdka} of the waves of the absorber particles. \vspace{2\baselineskip}

\section{Derivation IV}\label{absch:deriv4}
The \raisebox{4\baselineskip}[0pt][0pt]{\hypertarget{ta:deriv4}{}}parts of absorber theory, which have been reported up to now, have been named \mbox{derivation\,I} and \mbox{derivation\,II} by Wheeler and Feynman\!\cite{Wheeler:absorber}. In their section \mbox{derivation\,III} they discuss the relativistic generalization of the theory. We will not delve into that subject. 

\mbox{Derivation\,IV} is as well a part of the absorber theory of Wheeler and Feynman, which insofar would belong into our previous section. But actually \mbox{derivation\,IV} is exceptional, because it is the only part of absorber theory which does not depend on the strange postulate \eqref{ofdnjgkmfc} on the indices of refraction. For that reason the form of the result of \mbox{derivation\,IV} is --- as we will see immediately --- quite different from the form of the results which were presented in the previous section. 

In \mbox{derivation\,IV}, Wheeler and Feynman are considering a \al comp\-lete absorber\arp . That is a system of a finite number of charged point\bz{-}particles $q_k$, which are located in a finite space volume $V$ and are exerting forces upon one another due to electromagnetic interactions\bz{-}at\bz{-}a\bz{-}distance. But these particles don't radiate energy into the outer space (all radiation energy will be absorbed within the absorber), and no radiation energy is coming in from outer space. Therefore a test charge outside of the absorber volume $V$ would sense no force: 
\begin{align}
\sum _k\tfrac{1}{2}(F^{(k)}_r+F^{(k)}_a)=0\quad\text{(outside of }V) \label{osjsgvsygf}
\end{align}
$F^{(k)}_r$ and $F^{(k)}_a$ are the retarded and advanced electromagnetic forces exerted by the particle $q_k$. According to \eqref{osjsgvsygf}, outside of $V$ the forces $F^{(k)}_r$ and the forces $F^{(k)}_a$ must interfere destructively and mutually extinct completely. That is possible only, if 
\begin{align}
\sum _kF^{(k)}_r=\sum _kF^{(k)}_a=0\quad\text{(outside of }V) \label{ksdghsdgd}
\end{align}
holds. Consequently also 
\begin{align}
\sum _k\tfrac{1}{2}(F^{(k)}_r-F^{(k)}_a)=0\quad\text{(outside of }V)  \label{lsjhgdjhsxf}
\end{align}
must hold. In contrast to the fields \eqref{osjsgvsygf} and \eqref{ksdghsdgd}, the field \eqref{lsjhgdjhsxf} has no sources in $V$. Consequently 
\begin{align}
\sum _k\tfrac{1}{2}(F^{(k)}_r-F^{(k)}_a)=0\quad\text{(everywhere)}  \label{ksadgnbsew}
\end{align}
must hold. This statement may seem trivial at the first moment, but it allows for an important conclusion. According to action\bz{-}at\bz{-}a\bz{-}distance electrodynamics, this is the\pagebreak  
\begin{subequations}\begin{align}
&\text{force acting upon }q_j\text{ at space\bz{-}time point }x_j\, =\notag \\ 
&=F(x_j)=\sum _{k\neq j}\frac{1}{2}\Big( F^{(k)}_r(x_j)+F^{(k)}_a(x_j)\Big)\, =\label{ksdghsafda}\\ 
&=\sum _{k\neq j}F^{(k)}_r(x_j)+\underbrace{\frac{1}{2}\Big( F^{(j)}_r(x_j)-F^{(j)}_a(x_j)\Big)}_{F_{\text{rad}}}\, -\label{ksdghsafdb} \\ 
&\qquad -\underbrace{\sum _{k}\frac{1}{2}\Big( F^{(k)}_r(x_j)-F^{(k)}_a(x_j)\Big)}_{\stackrel{\eqref{ksadgnbsew}}{=}0}\ .\notag 
\end{align}\end{subequations}
The last term is zero according to \eqref{ksadgnbsew}. The first term is describing the force which according to Maxwell's electrodynamics is accelerating the particle $q_j$. Therefore the second term must be the 
\vspace{-3ex}\begin{align}
&\text{radiation back-reaction}=\tfrac{1}{2}(F^{(j)}_r-F^{(j)}_a)\stackrel{\eqref{ksdgvsgc}}{=}F_{\text{rad}}\ .\label{isayhgdrfd} 
\end{align} 
According to the premises of the model, each particle is exerting forces onto other particles only, but not onto itself. Therefore the sum in \eqref{ksdghsafda} is running over $k\neq j$ only. In the radiation back\bz{-}reaction, however, surprisingly forces $F^{(j)}$ are showing up, which the particle $q_j$ is exerting upon itself. The term $F_{\text{rad}}$ is different from zero, because the expression \eqref{ksadgnbsew} is equal to zero. The property \eqref{ksadgnbsew} of a complete absorber, which at first sight seems so unimpressive, brings about the intricate effect, that a particle is able to act indirectly upon itself after all\ARZ  Wheeler and Feynman comment:\vspace{-.2ex}\enlargethispage{2\baselineskip} 
\begin{flushright}\parbox{.88\linewidth}{\al we have shown that the half\bz{-}advanced, half\bz{-}retarded fields of the theory of action at a distance lead to a satisfactory account of the mechanism of radiative reaction and to a\linebreak}\parbox{.09\linewidth}{\vspace{-2ex}\begin{align}\label{osdghfsds}\end{align}}\end{flushright} 
\begin{flushright}\parbox{.88\linewidth}{description of the action of one particle on another in which no evidence of the advanced fields is apparent. We find in the case of an absorbing universe a complete equivalence between the theory of Schwarzschild and Fokker on the one hand and the usual formalism of electrodynamics on the other.\ar}\parbox{.09\linewidth}{\begin{align}\notag\end{align}}\end{flushright} 

\noindent Wheeler und Feynman are pointing out, that Dirac\!\cite{Dirac:classelectron} already in 1938 indicated the expression \eqref{isayhgdrfd} --- besides a factor 1/2 due to differing definitions --- for the radiation back\bz{-}reaction, see Dirac's equation (11). But Dirac arrived at this result due to purely mathematical tinkering, while Wheeler and Feynman claim to have derived it from the physical model of a complete absorber. 

As a complete absorber they mention \al an absorbing universe\arp . From today's point of view one might remark, that the ubiquitous cosmic background radiation probably does exclude any smaller system than the whole universe as a candidate for a complete absorber. By 1945, when Wheeler and Feynman published their absorber theory, that background radiation was still unknown. Anyway a look through the mount Wilson telescopes into the depths of intergalactic space probably even by then left not much hope that a smaller part of the universe could be a complete absorber. Then it must be asked, however, what reasonable meaning could be assigned to the specification \al (outside of $V$)\ar  in the equations \eqref{osjsgvsygf} through \eqref{ksadgnbsew}. This derivation of \eqref{ksadgnbsew} certainly is not built on firmer physical grounds than Dirac's mathematical tinkering. 

Probably it would be better, not to \al derive\ar \eqref{ksadgnbsew}, but instead to postulate \eqref{ksadgnbsew} as the \emph{defining property} of a complete absorber. But then of course the immediate consequence \eqref{isayhgdrfd} would almost become a part of the definition. From this point of view, the result \eqref{isayhgdrfd} can hardly be considered to mark some really significant improvement over Dirac's findings. 

\section{Applying the reasonable assumption (15b)}\label{absch:problem1}
The \raisebox{4\baselineskip}[0pt][0pt]{\hypertarget{ta:problem1}{}}derivations of the radiation back\bz{-}reaction formula in \eqref{kdjhsddgsg} and in \eqref{sjgksdghsfg}, and also the computation \eqref{klsnjhgdka} of the advanced fields nearby the source, all are based on the 
\begin{align}
\text{\bfseries Postulate:}\left\{\mbox{\parbox{.72\linewidth}{$n\neq 1$ applies for retarded interactions.\newline{}$n=1$ applies for advanced interactions.}}\right.\hspace{-3em}\tag{\ref{ofdnjgkmfc}}
\end{align}
In the next section we will present arguments, why assumption \eqref{ofdnjgkmfb}, saying that the same dielectric constant should be applied to both retarded and advanced interactions, seems much more reasonable, and why the postulate \eqref{ofdnjgkmfc} is not at all justified. If the computations are based on \eqref{ofdnjgkmfb}, then the sum of the advanced fields of the absorber at the position of the source $q$ becomes, in case that the absorber is a plasma ($n<1$), 
\begin{align}
\widetilde{\boldsymbol{E}}_a^{\text{total}}(\omega ,\boldsymbol{r})\ggstackrel{\eqref{kdjhsddgsg}}-\widetilde{\boldsymbol{\dot{v}}}(\omega )\,\frac{e^2qN}{2m_ec^4(4\pi\epsilon _0)^2}\,\frac{8\pi}{3}\!\underbrace{\inte _0^\infty\!\dif R\,\exp\{ -\omega 2R\kappa /c\}}_{c/(2\omega\kappa )}\notag\\ 
&=-\frac{2qe^2N\pi}{3c^3(4\pi\epsilon _0)^2m_e\omega\kappa }\,\widetilde{\boldsymbol{\dot{v}}}(\omega )\notag\\ 
\ggstackrel[.2]{\eqref{lsdghjsavc}}-\frac{ie^2N}{4\epsilon _0m_e\omega ^2\kappa }\cdot\frac{2q}{3c^34\pi\epsilon _0}\,\widetilde{\boldsymbol{\ddot{v}}}(\omega )\ . 
\end{align} 
This result differs from the respective result \eqref{kdjhsddgsg} of Wheeler and Feynman by the first fraction on the right side. As it is reasonable to assume that the extinction is proportional to the plasma's density, $\kappa\sim N$, the total advanced field does not significantly depend on the plasma's properties. But different from the classical radiation back\bz{-}reaction \eqref{ksdgvsgc}, here $\widetilde{\boldsymbol{E}}_a^{\text{total}}$ lags the phase of $\widetilde{\boldsymbol{\dot{v}}}$ by $\pi $, and consequently lags the phase of $\widetilde{\boldsymbol{\ddot{v}}}$ by $3\pi /2$. The damping integral is resulting into a phase shift of $\pi $ versus the single parts \eqref{oasaghdews} of $\boldsymbol{E}_a^{\text{total}}$. 

If the absorber is a material with bound electrons ($n>1$), then the plausible assumption \eqref{ofdnjgkmfb} leads to the result 
\begin{subequations}\label{osjhgksgs}\begin{align}
&\widetilde{\boldsymbol{E}}_a^{\text{total}}(\omega ,\boldsymbol{r})\stackrel{\eqref{posahnbgkwr}}{=}-\widetilde{\boldsymbol{\dot{v}}}\,\frac{Ne^2p}{\epsilon _0m_e\omega ^2}\,\frac{q\omega ^2}{2c^4 4^2\pi ^2\epsilon _0}\,\frac{8\pi}{3}\,\underbrace{\frac{2}{1+n-i\kappa}}_{\approx 1}\,\cdot\notag\\ 
&\qquad\cdot\underbrace{\frac{2}{1+n+i\kappa}}_{\approx 1}\;\underbrace{\inte _{R_{\text{cav}}}^{\infty}\!\dif R\,\exp\Big\{ -\,\frac{2(R-R_{\text{cav}})\,\omega\,\kappa}{c}\Big\}}_{c/(2\omega\kappa )} \notag\\ 
&=-\frac{2}{1+n-i\kappa}\,\frac{2}{1+n+i\kappa}\,\frac{Ne^2p}{4\epsilon _0m_e\omega \kappa }\,\frac{2q}{3c^34\pi\epsilon _0}\,\widetilde{\boldsymbol{\dot{v}}}\notag\\ 
&=G\cdot \frac{2q}{3c^34\pi\epsilon _0}\,\widetilde{\boldsymbol{\ddot{v}}}\\ 
G&\equiv -\frac{2}{1+n-i\kappa}\,\frac{2}{1+n+i\kappa}\,\frac{iNe^2p}{4\epsilon _0m_e\omega ^2\kappa } \ .\label{osjhgksgsb}
\end{align}\end{subequations} 
Here we introduced the name $G$ for the factor, by which this result differs from the respective result \eqref{sjgksdghsfg} of Wheeler and Feynman. It doesn't make much difference whether or not we keep the somewhat dubious trick with the transmission coefficient \eqref{ksksdhndgg}. (Note that we inserted $\kappa $ with different signs for retarded versus advanced fields in both the transmission coefficients and in the exponent.) In any case $\widetilde{\boldsymbol{E}}_a^{\text{total}}$ is essentially in phase with $+\widetilde{\boldsymbol{\dot{v}}}$, because the material factor $p(\omega )$ in the refractive index \eqref{ppfbjndljbhds} must essentially be $<0$, to get $n>1$. Thus it's phase differs by $\pi $ from the phase factor in case $n<1$. We have emphasized above, that the imaginary part $i\kappa$ of the refractive index may be significant; but it would be rather flimsy now to postulate it's value such that the desired phase relation $\widetilde{\boldsymbol{E}}_a^{\text{total}}\!\sim\widetilde{\boldsymbol{\ddot{v}}}$ is achieved. As $p$ is not canceled,  $\widetilde{\boldsymbol{E}}_a^{\text{total}}$ is depending on the absorber's material. 

Obviously there is no chance to derive from absorber theory the classical radiation back\bz{-}reaction, as long as we insist on assumption \eqref{ofdnjgkmfb}. This assumption actually results in even worse consequences, if we compute the value of $\widetilde{\boldsymbol{E}}_a^{\text{total}}$ at some spot $\boldsymbol{r}+\boldsymbol{s}$ in the absorber cavity of figure~\ref{fig:kavitaet} nearby the source $q$, \ggie $s\equiv|\boldsymbol{s}|\ll R_{\text{cav}}$ and $n>1$. As explained at \eqref{klshgsdgvsdb}, we get the result 
\begin{align}
&\widetilde{\boldsymbol{E}}_a^{\text{total}}(\omega ,\boldsymbol{r}+\boldsymbol{s})\stackrel{\eqref{osjhgksgs},\eqref{lsdghjsavc}}{=}-i\omega G\,\frac{2q}{3c^3 4\pi\epsilon _0}\,\widetilde{\boldsymbol{\dot{v}}}(\omega )\cdot I(\omega ,\boldsymbol{r}+\boldsymbol{s}) \notag\displaybreak[1]\\ 
&I\equiv\frac{3}{8\pi}\inte _0^{2\pi}\!\dif\varphi\inte _0^\pi\!\dif\vartheta\,\sin ^3\vartheta\,\exp\Big\{ -i\omega\,\frac{s\cos (\boldsymbol{s},\boldsymbol{n})}{c}\Big\}\ .   
\end{align} 
We encountered the same expression for $I$ already in \eqref{klshgsdgvsdb}, and have indicated the solution of this integral in \eqref{ksksghdgfs}. If $s$ exceeds some few wavelengths, then 
\begin{align}
I\stackrel{\eqref{osajnhgds}}{\approx}\frac{3\sin ^2(\boldsymbol{s},\boldsymbol{\dot{v}})}{2} \,\frac{c}{\omega s}\,\frac{1}{2i}\,\Big[\,\exp\{ +i\omega s/c\} -\exp\{ -i\omega s/c\}\Big] 
\end{align}
holds, which leads to 
\begin{align}
\widetilde{\boldsymbol{E}}_a^{\text{total}}(\omega ,\boldsymbol{r}+\boldsymbol{s})&=-G\,\frac{\sin ^2(\boldsymbol{s},\boldsymbol{\dot{v}})}{2}\,\frac{q}{sc^24\pi\epsilon _0}\,\widetilde{\boldsymbol{\dot{v}}}(\omega )\,\cdot \notag\\ 
&\quad\cdot \Big[\,\exp\{ +i\omega s/c\} -\exp\{ -i\omega s/c\}\Big]\ .\label{aiasgnsdy}
\end{align}
This result again differs from \eqref{klsnjhgdka} by the factor $G=\eqref{osjhgksgsb}$. In this case, the more plausible assumption \eqref{ofdnjgkmfb} results into observable consequences: The result \eqref{klsnjhgdka} of the sum of the advanced fields was by $\pi $ out of phase with the advanced field \eqref{klsnjhgdkb} radiated  by $q$ itself. We have emphasized, that the destructive interference is necessary, because no experimenter ever observed an advanced field. 

\eqref{aiasgnsdy} however is by about $\pi /2$ out of phase with \eqref{klsnjhgdkb}, because $G$ is essentially imaginary (though it's real part may not be negligible). Thus there is no complete destructive interference. The advanced fields \eqref{aiasgnsdy} should be observable, but they are not observed. That means that absorber theory would be incompatible with experimental evidence, if it would be based onto the plausible assumption \eqref{ofdnjgkmfb}. 

\section{The perplexing postulate (15c)}\label{absch:why}
We \raisebox{4\baselineskip}[0pt][0pt]{\hypertarget{ta:why}{}}emphasized already several times, that \eqref{ofdnjgkmfb} seems much more plausible to us than the postulate \eqref{ofdnjgkmfc}. But why? Are there convincing arguments for \eqref{ofdnjgkmfb}? Are there convincing arguments against \eqref{ofdnjgkmfc}? 

Lets assume the system under consideration to be composed of \ggeg  $10^{24}$ electrically charged particles. We appoint one of them the source $q$, all other particles $q_k$ constitute the absorber. A perfect description of this system in terms of classical action\bz{-}at\bz{-}a\bz{-}distance electrodynamics would require the computation of the retarded interactions and the advanced interactions inbetween all 
\centerline{$10^{24}(10^{24}-1)/2\approx 5\cdot 10^{47}$} pairs of particles. That of course exceeds by far our capabilities. If we want to achieve some result at all, then we cannot avoid to simplify the model appreciably. Wheeler and Feynman do that --- as is common practice in classical electrodynamics --- by restricting the computation to the source and one absorber particle only, and allow for interactions with all other particles, and interactions of those other particles with again other particles\,\dots\ by insertion of an index of refraction $n$, due to which the \al dielectric background\ar  is overall considered. 

Thus far their computation is reasonable and regular. But it is perplexing, that Wheeler and Feynman now decide for an index of refraction $n\neq 1$ valid for the retarded interaction between the source and some certain absorber particle, but for the advanced interaction between the source and \emph{the same} absorber particle they choose the index of refraction $n=1$. We will try to explicate two times, why that assumption is completely baffling. First in a more pictorial manner, and then in a more technical manner. 

For the pictorial explanation, lets consider again the bottom sketch in figure~\ref{fig:wellen} \vpageref[\negmedspace ]{fig:wellen}. There we see the superposition of eight red\bz{-}painted advanced fields collapsing onto eight absorber particles, and one blue\bz{-}painted retarded field spreading out from $q$. In a more realistic sketch we would see the superposition not only of eight, but of a huge number of red\bz{-}painted advanced fields collapsing onto the same number of absorber particles, and one blue\bz{-}painted retarded field spreading off from $q$. Furthermore we know, that the red fields contribute \mbox{50\,\%} to the total field amplitude, and the blue field is contributing the remaining \mbox{50\,\%}. 

It is clear that in the framework of action\bz{-}at\bz{-}a\bz{-}distance electrodynamics, fields only have the status of computation aids on the theorist's paper. Each particle in the system exerts upon each other particle in the system a retarded and an advanced action\bz{-}at\bz{-}a\bz{-}distance force. When we say that the sum of the amplitudes of the red fields and the amplitude of the blue field are equal at the location of some certain particle $q_j$, then that is just a symbolic manner of speaking for the statement, that the sum of the advanced forces exerted by all absorber particles $q_k$ upon the particle $q_j$ is equal to the retarded force exerted by the source $q$ upon $q_j$. 

All these forces are acting at the same time upon $q_j$. The essential point is: $q_j$ senses just one total force. It does not decompose this force into blue and red forces, nor into retarded and advanced forces. Only we are doing that in our theoretical model. $q_j$ is not sensing a retarded force which is accelerating it by $\boldsymbol{\dot{v}}_{j,r}$, and an advanced force which is accelerating it by $\boldsymbol{\dot{v}}_{j,a}\neq\boldsymbol{\dot{v}}_{j,r}$. Instead $q_j$ is sensing \emph{exactly one} total force. This total force is accelerating $q_j$ by $\boldsymbol{\dot{v}}_{j}$. Caused by this acceleration, $q_j$ now exerts on it's part action\bz{-}at\bz{-}a\bz{-}distance forces onto all other particles in the system, and thereby brings about an indirect, secondary interaction between the source and the absorber particles $q_k$. 

We don't compute the forces between $q_j$ and the other particles explicitly, but we subsume these forces into the index of refraction $n$. A particle, that is conveying the secondary interactions between the source and the absorber particles, does not react in different manners onto the retarded and the advanced fields, but it reacts in exactly one unique manner onto the sum of the forces to which it is subjected. The same holds true for any other particle in the system. The overall combination of the secondary forces to an index of refraction can consequently result only into a unique index of refraction, which is valid for both retarded and advanced interactions. 

The more technical consideration is motivated by the invariance of action\bz{-}at\bz{-}a\bz{-}distance electrodynamics under time inversion. As is well known, Maxwell's equations are invariant under time inversion. This symmetry is usually abandoned, if Maxwell's theory is applied to the description of radiation processes. The retarded radiation fields \eqref{lksjngfdkf} with index $_r$, which are radiated by an accelerated charge, form spherical waves spreading off from the radiating charge. The advanced radiation fields \eqref{lksjngfdkf} with index $_a$ form spherical waves collapsing onto the accelerated charge. The latter is never observed. Therefore usually the advanced solutions are skipped, and symmetry under time inversion is lost. 

In contrast, action\bz{-}at\bz{-}a\bz{-}distance electrodynamics postulates, that any interaction between charged particles can and must be described as the superposition of a retarded and an advanced interaction:\pagebreak  
\begin{align}
\underbrace{F_r}_{\displaystyle\hspace{-2em}\text{Maxwell}\hspace{-2em}}\quad\longleftrightarrow\quad\underbrace{\tfrac{1}{2}(F_r+F_a)}_{\displaystyle\hspace{-2em}\text{action-at-a-distance}\hspace{-4em}}
\end{align}
Under time inversion, retarded interactions become advanced interactions, and vice versa. Therefore action\bz{-}at\bz{-}a\bz{-}distance electrodynamics is by construction invariant under time inversion, even in case of radiation phenomena. 

Wheeler and Feynman dismiss this invariance due to postulate \eqref{ofdnjgkmfc}. In their absorber theory, the retarded interaction is not equal to the time\bz{-}inverted advanced interaction; instead the retarded interaction is equal to the time\bz{-}inverted advanced interaction multiplied by $n\neq 1$. Wheeler and Feynman invoke action\bz{-}at\bz{-}a\bz{-}distance electrodynamics at many places in their article. But it is just the \emph{deviation} from action\bz{-}at\bz{-}a\bz{-}distance electrodynamics, which enabled them to arrive at the results \eqref{kdjhsddgsg}, \eqref{sjgksdghsfg}, and \eqref{klsnjhgdka}. We proved in the previous section, that postulate \eqref{ofdnjgkmfc}, by which the time\bz{-}inversion symmetry of retarded and advanced interactions is abandoned, is an indispensable pre\bz{-}condition for all those results. Lets hear, how Wheeler and Feynman justified their postulate \eqref{ofdnjgkmfc}. On page~161 bottom, right column, they write: 
\begin{flushright}\parbox{.88\linewidth}{\al The advanced force acting on the source due to the motion of a typical particle of the absorber is an elementary interaction between two charges, propagated with the speed of light in vacuum. On the other hand, the disturbance which travels outward from the source and determines the motion of the particle in question is made up not only of the proper field of the originally accelerated charge, but also of the secondary fields generated in the material of the absorber. The elementary interactions are of course propagated with the speed of light; but the combined disturbance travels, as is well known from the theory of the refractive index, at a different speed, $c/(\text{refractive index})=c/n$.\ar }\parbox{.09\linewidth}{\begin{align} \end{align}}\end{flushright} 
This is merely the assertion, which we encoded formally in \eqref{ofdnjgkmfc}; but it is no justification. Not any argument is put forward for the perplexing decision, to abandon at this point the time\bz{-}inversion symmetry of action\bz{-}at\bz{-}a\bz{-}distance electrodynamics. 

A substantially better explanation for the asymmetric indices of refraction can be found in a lecture\!\cite{Feynman:nobel}, which Feynman gave twenty years after the publication of absorber theory. He recalls in that lecture, that he struggled quite a long time with unsatisfactory results, which probably looked similar to those, which we have computed in the previous section: material\bz{-}dependent, frequency\bz{-}dependent, wrong phases, in short: a mess, Feynman got stuck. Thus he went to Wheeler, to discuss the problem: 
\begin{flushright}\parbox{.88\linewidth}{\al Professor Wheeler used advanced waves to get the reaction back at the right time and then he suggested this: If there were lots of electrons in the absorber, there would be an index of refraction $n$, so, the retarded waves coming from the source would have their wave lengths slightly modified in going through the absorber. Now, if we shall assume that the advanced waves come back from the absorber without an index --- why? I don't know, let's assume they come back without an index [\dots ] And when we estimated it, [\dots ] sure enough, it came out that the action back at the source was completely independent of the properties of the charges that were in the surrounding absorber. Further, it was of just the right character to represent radiation resistance\arp .}\parbox{.09\linewidth}{\begin{align} \end{align}}\end{flushright} 
Thank you, Feynman, and compliment, for these clear words. Thus they needed to arrive somehow at the desired results, and that need prompted postulate \eqref{ofdnjgkmfc}. Such practice is acceptable, in principle. Consider for example Bohr's model of the atom, published in 1913\!\cite{Bohr:modell1913}. Bohr strove for a model, which first should be compatible with Rutherford's experimental observations of $\alpha $\bz{-}particle scattering, which second should reproduce as exactly as possible the observed absorption\bz{-} and emission spectra, and which third should not deviate more than unavoidable from classical mechanics and classical electrodynamics. \al Not deviate more than unavoidable from\ar  clearly does not mean \al comply with\arp . Bohr was convinced, that the results of atomic physics were beyond the scope of the classical theories. Thus he took the liberty to declare some postulates, which clearly could not be reconciled with the established classical theories. 

Wheeler and Feynman discovered that a beautiful and elegant theory of radiation back\bz{-}reaction can be constructed if only one declares the postulate \eqref{ofdnjgkmfc}, which clearly can not be reconciled with action\bz{-}at\bz{-}a\bz{-}distance electrodynamics (nor with Maxwell's electrodynamics). Insofar the analogy to Bohr's model seems to fit. But there still are objections. 

Bohr's model had a high prognostic power. Through the following decade, more and more spectra of atoms and molecules, and more and more details within these spectra, were interpreted and classified by means of Bohr's model. And the model itself was constantly further improved, refined, and adapted to new experimental findings. 

The prognostic power of absorber theory, however, is rather poor. Essentially the balance is: 1 postulate, \ggie \eqref{ofdnjgkmfc}, for the derivation of 1 result, \ggie the classical radiation back\bz{-}reaction \eqref{ksdgvsgc}. Furthermore \eqref{ofdnjgkmfc} leads to the classical radiation back\bz{-}reaction only in case of stationary radiation. In case of time\bz{-}dependent pulsed radiation, it leads to senseless results. Consider for example a source $q$ and an absorber particle $q_k$, which are connected by a\linebreak\centerline{\begin{overpic}[trim=0 1mm 0 -1mm]{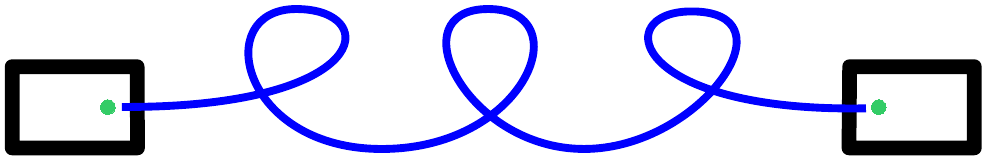}
\put(7.5,3.5){\textcolor{dklgrn}{$q$}}
\put(91,3.5){\textcolor{dklgrn}{$q_k$}}
\end{overpic}}\vspace{2ex}\newline{}glass fiber cable, as commonly used in telecommunication. Let the length of the glass fiber be \mbox{$200\,\text{m}$}, and let it's index of refraction be $n\approx 1.5$\,. Let the source emit retarded digital signals at a rate of \mbox{$1\,\text{Gbit}/s$}. Thus the length of one bit is $10^{-9}\text{s}$, and it arrives $10^{-6}\text{s}$ after the emission at the absorber particle $q_k$. According to classical action\bz{-}at\bz{-}a\bz{-}distance electrodynamics, $q_k$ emits advanced signals, which propagate opposed to the usual direction of time, have a length of $10^{-9}\text{s}$, and arrive $10^{-6}\text{s}$ before their emission at the source $q$. Viewed through the glasses of usual time direction, the advanced signals are propagating from $q$ to $q_k$, and they are at any time and everywhere in phase with the retarded signals emitted by $q$. 

According to postulate \eqref{ofdnjgkmfc} of absorber theory, however, only the retarded signals would follow all the loops of the fiber at a speed of $2\cdot 10^8\text{m/s}$. The advanced signals would choose the shortest path between $q$ and $q_k$, would run along this path at a speed of $3\cdot 10^8\text{m/s}$, and would ignore the fiber's loops. Any telecommunication net would break down immediately, if postulate \eqref{ofdnjgkmfc} would be correct.\vspace{1\baselineskip} 

Of course we are not the first to criticize the strange postulate \eqref{ofdnjgkmfc}. In 1970, Kamat\!\cite{Kamat:absorber} considered Wheeler\bz{-}Feynman absorber theory. He announced that he would give a derivation of radiation back\bz{-}reaction along the lines proposed by Wheeler and Feynman, but deviating from Wheeler and Feynman in point (iv) of below list of steps, which Kamat compiled in section~2 of his article. There Kamat is using the wording \al net\ar  interaction for an interaction which is the superposition of of the \al elementary\ar  interaction between source $a$ and absorber particle $b$ and the indirect interactions mediated via all other absorber particles: 
\begin{ggitemize}
\ggitem[\al (i)]{Let the source particle $a$ receive the acceleration at time $t$.} 
\ggitem[(ii)]{It radiates a fully retarded electromagnetic disturbance, which travels outwards.} 
\ggitem[(iii)]{The net retarded field disturbs the absorber particle $b$.} 
\ggitem[(iv)]{The absorber particle interacts back on the source through a fully advanced field, which is also a net field, hence not elementary.} 
\ggitem[(v)]{Summing over all absorber particles $b\neq a$, the radiation reaction field is calculated.\ar } 
\end{ggitemize} 
To justify (iv), Kamat added: \al The given absorber particle $b$ receives the net retarded field which is the superposition of the proper field of the source $a$ and those of the absorber particles other than $b$. By the principle of action and reaction, the response field of $b$ should interact back with the particles other than $b$ and the net field should reach the source $a$. This means that we should include the refractive index for the returned response field as well.\ar  

We completely agree to the proposal stated by Kamat. If he had proceeded along this proposal, he would of course have arrived at the solutions which we presented above in section~\ref{absch:problem1}. But unfortunately he did not\ARZ  Instead Kamat\!\cite{Kamat:absorber} proceeded like this: In the equation for the acceleration of the source particle $q$ due to the advanced reaction from the absorber particle $q_k$ 
\begin{align}
q\boldsymbol{E}_a(t_r,\boldsymbol{r})\stackrel{\eqref{ofdnjgkmfb}}{=}q\inte _{-\infty}^{+\infty}\!\frac{\dif\omega}{2\pi}\,\widetilde{\boldsymbol{E}}_a(\omega ,\boldsymbol{x})\,\exp\{ -i\omega (t_r\underbrace{+Rn/c-Rn/c}_{\text{phase-shift}})\} \label{klsdgnhays}
\end{align}
he introduced a summation over slices of frequencies, the thickness of each slice being $2\epsilon _l$: 
\begin{align}
q\boldsymbol{E}_a(t_r,\boldsymbol{r})&=\sum _lq\inte _{\omega _l-\epsilon _l}^{\omega _l+\epsilon _l}\!\frac{\dif\omega}{2\pi}\,\widetilde{\boldsymbol{E}}_a(\omega ,\boldsymbol{x})\,\cdot\notag\\ 
&\quad\cdot\exp\{ -i(\omega _lt_r+R\underbrace{\omega _ln/c}_{\kappa _l}-R\underbrace{\omega _ln/c}_{\kappa _l})\} \label{lsdafghjn} 
\end{align}
Next Kamat proved in his three equations (8) the relation 
\begin{subequations}\begin{align}
\kappa _l&=\frac{\omega }{u_l}+\frac{\omega _l^2}{v_l^2}\,\difq{v}{\omega}\bigg| _{\omega _l}\quad\text{for }\omega _l-\epsilon _l\leq\omega\leq\omega _l+\epsilon _l\label{oweglsrga}\\ 
u_l&\equiv\text{group velocity nearby }\omega _l\notag\\ 
v_l&\equiv\text{phase velocity nearby }\omega _l\ ,\notag 
\end{align}
which is an excellent approximation provided that $\epsilon _l$ is chosen sufficiently small. Even the less accurate approximation 
\begin{align}
\kappa _l\approx\frac{\omega }{u_l}\quad\text{for }\omega _l-\epsilon _l\leq\omega\leq\omega _l+\epsilon _l\label{oweglsrgb} 
\end{align}\end{subequations}
is acceptable for very small $\epsilon _l$. This expansion of $\kappa $ clearly can not change the fact, that the phase shift marked in the exponent of \eqref{klsdgnhays} is zero. But as Kamat did not want to accept this result, he procured a phase shift which is different from zero due to the following strange decision:  
\begin{subequations}\label{klsghnsgs}\begin{align}
\eqref{oweglsrga}&\text{ applies for advanced interactions}\\ 
\eqref{oweglsrgb}&\text{ applies for retarded interactions} 
\end{align}\end{subequations}
That of course is nothing but the postulate \eqref{ofdnjgkmfc} of Wheeler and Feynman in a new disguise. Due to his postulate \eqref{klsghnsgs}, which he did not state explicitly, but which he applied implicitly, Kamat arrived at his equation (10):  
\begin{align}
q\boldsymbol{E}_a(t_r,\boldsymbol{r})&\stackrel{\eqref{lsdafghjn},\eqref{klsghnsgs}}{=}\sum _lq\inte _{\omega _l-\epsilon _l}^{\omega _l+\epsilon _l}\!\frac{\dif\omega}{2\pi}\,\widetilde{\boldsymbol{E}}_a(\omega ,\boldsymbol{x})\,\cdot\notag\\ 
&\quad\cdot\exp\Big\{ -i\Big(\omega _lt_r-R\frac{\omega _l^2}{v_l^2}\,\difq{v}{\omega}\bigg| _{\omega _l}\Big)\Big\} \label{ksdfgnhb}
\end{align} 
Kamat gives no justification or explanation at all for \eqref{klsghnsgs}. Obviously his derivation is inconsistent with his own premise (iv) cited above. Thus the demonstration, that the classical radiation back\bz{-}reaction can be derived from \eqref{ksdfgnhb}, is no advancement over the derivation given by Wheeler and Feynman, and all our objections against postulate \eqref{ofdnjgkmfc} are valid objections against \eqref{klsghnsgs} as well. 

\section{Correcting the interpretation of (42a)}\label{absch:error19}
In \raisebox{4\baselineskip}[0pt][0pt]{\hypertarget{ta:error19}{}}\eqref{lksakdsgjc} we cited Wheeler and Feynman with the words \al that the advanced field of the absorber is equal in the neighborhood of the accelerated particle to the difference between half the retarded field and half the advanced field which one calculates for the source itself\arp . This assertion is not correct. Only the projections of the fields onto the axis of $\boldsymbol{\dot{v}}$ are equal: \eqref{klsnjhgdka}, which is equivalent to the result\!\cite[equation\,(19)]{Wheeler:absorber} in the publication of Wheeler and Feynman, is equal to \eqref{klsnjhgdkc}, but not equal to \eqref{klsnjhgdkb}\ARZ The fields \eqref{klsnjhgdkb} radiated by the accelerated source $q$ are spherical waves. Their electric and magnetic fields are at any point perpendicular to the straight line between the source and that point. Have again a look at figure\,\ref{fig:geschwubeschl} on page\,\pageref{fig:geschwubeschl}: The fields \eqref{klsnjhgdkb} very well have components, which are perpendicular to $\boldsymbol{\dot{v}}$. In contrast, in the sum \eqref{sjgksdghsfg} of the advanced fields radiated by all absorber particles, indeed the components perpendicular to $\boldsymbol{\dot{v}}$ mutually compensate nearby the source, and only the components parallel to $\boldsymbol{\dot{v}}$ survive. 

The consequences can be proved by experiment. The superposition of the advanced fields emitted by the absorber particles, which lag by $\pi $ versus the phase of the advanced fields emitted by the source $q$ (upper sketch in fig.\,\ref{fig:wellen} on page\,\pageref{fig:wellen}), only compensate those components of this field, which are parallel to $\boldsymbol{\dot{v}}$. The components of the advanced field emitted by the source $q$, which are vertical to $\boldsymbol{\dot{v}}$, are not compensated by the advanced fields emitted by the absorber particles, and consequently should be measurable. But they are not observed. Thus absorber theory is refuted by experiment.   

Equivalent conclusions must be drawn with regard to the retarded field emitted by the source $q$: Only it's component parallel to $\boldsymbol{\dot{v}}$ is constructively enforced by the advanced fields of the absorber particles (bottom sketch in fig.\,\ref{fig:wellen}), but not it's components vertical to $\boldsymbol{\dot{v}}$. Consequently the observed total field can not consist --- as assumed by absorber theory --- of \mbox{50\,\%} contributed by the superposition of the advanced fields of the absorber particles, and \mbox{50\,\%} contributed by the retarded field of the source. 

Thus absorber theory is incompatible with experimental evidence, no matter whether the strange postulate \eqref{ofdnjgkmfc} is accepted or whether the reasonable assumption \eqref{ofdnjgkmfb} is preferred. Still absorber theory should not be dismissed over\bz{-}hastily. We will forward arguments in the next section for the conjecture, that a quantum\bz{-}theoretical formulation of absorber theory probably would be free of the flaws of classical absorber theory, on which we have dwelt in this and the previous section. 

\section{Quantized absorber theory?}\label{absch:altern}
We \raisebox{4\baselineskip}[0pt][0pt]{\hypertarget{ta:altern}{}}will keep the discussion of the quantum\bz{-}theoretical perspectives of absorber theory very short and give only some indications, because this article is dedicated to the absorber theory of Wheeler und Feynman. These authors designed absorber theory as a purely classical theory. 

We know that neither Maxwell's classical electrodynamics nor classical action\bz{-}at\bz{-}a\bz{-}distance electrodynamics are able to describe the emission and absorption of electromagnetic radiation correctly (\ggie in accord with observation). Accelerated charges don't emit energy as spherical waves, but as photons. And the energy of a photon is absorbed by exactly one absorber particle. In their absorber theory, Wheeler and Feynman assume that any of a huge number of absorber particles can absorb an arbitrarily small part of the energy radiated by the source. If they thereby arrive at results, which can not be reconciled with experimental observation, then that is not really a surprise. It is just indicating, that they exceeded the limits, within which classical electrodynamics give correct results. 

If we assume in a quantum\bz{-}theoretical formulation of the theory, that the absorber particle sends an advanced photon back to the source, then an experimenter, who observes\footnote{Since some time, photons can be observed without absorption, see \ggeg\hspace{-.2em}\cite{Reiser:photdet}.} this half retarded, half advanced photon at some certain point in space\bz{-}time, will see --- provided that the resolution of the instruments is sufficient --- only this one photon, and no interferences with any other fields. 

In the quantum absorber theory, the retarded and the advanced half\bz{-}photon would of course be in phase everywhere and at any time. The notion of an \al orbit\ar  or \al path\ar  $\boldsymbol{r}(t)$ of an electron (or any other elementary particle) must be abandoned, as Heisenberg\!\cite{Heisenberg:umdeutung} emphasized in his seminal article on quantum mechanics. Consequently the time derivatives $\boldsymbol{\dot{v}}$ and $\boldsymbol{\dot{v}}_k$ (and the directions of these vectors) are not defined in quantum absorber theory. The electric and magnetic fields of the photons are oriented perpendicular to the direction of their propagation\cite[chap.\,15]{Gruendler:fieldtheory}. The issue regarding the projection of the fields onto the $\boldsymbol{\dot{v}}$\bz{-}axis thus is not solved by quantum absorber theory, but it becomes obsolete.   

By 1980 Cramer\!\cite{Cramer:absorbtheor} published a semi\bz{-}classical model, in which the exchange of a photon between a source and an absorber particle is described as follows:
\begin{ggitemize}
\ggitem{A radiating source emits a retarded field into a certain direction, and an advanced field of same strength into the opposite direction.}
\ggitem{The phase difference between the retarded and the advanced field at the position of the source is $\pi $.}
\ggitem{No spherical waves are emitted; instead the radiation has some certain direction $\boldsymbol{n}$. Therefore the amplitudes of the fields do not decrease with $R^{-1}$ as the radiation fields \eqref{lksjngfdkf} of Maxwell's theory; instead they are independent of the distance from the source.}
\end{ggitemize}
\hspace*{1em}\begin{overpic}[width=0.95\textwidth ,height=14mm, trim=0 -1mm 0 -1mm]{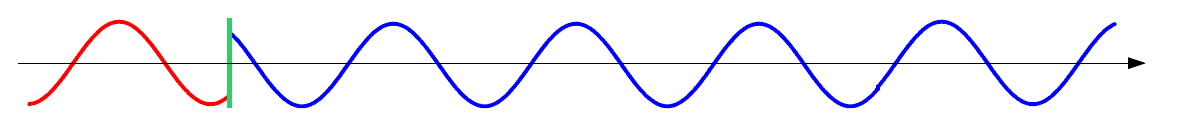}
\put(18.5,0){\textcolor{dklgrn}{$q$}}
\put(95,8){$ct$}
\end{overpic}
\begin{ggitemize}
\ggitem{An absorber particle absorbs an incoming field as follows: It radiates a retarded field of same strength, which is out of phase with the incoming retarded field by $\pi $, into the same direction $\boldsymbol{n}$; and it radiates an advanced field, which is in phase with the incoming field and directed opposite to the incoming field.}
\end{ggitemize}
\hspace*{1em}\begin{overpic}[width=0.95\textwidth ,height=14mm, trim=0 -1mm 0 -1mm]{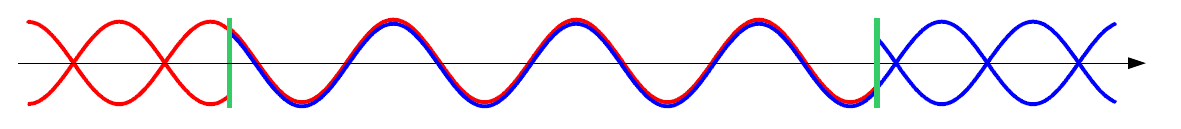}
\put(18.5,0){\textcolor{dklgrn}{$q$}}
\put(73,0){\textcolor{dklgrn}{$q_k$}}
\put(95,8){$ct$}
\end{overpic}\newline{}
\enlargethispage{2\baselineskip}Only in the space between $q$ and $q_k$, and only in the time interval between $t_r$ and $t_k$, the amplitude of the fields is different from zero, while outside this interval the fields completely annihilate due to destructive interference. The retarded and the advanced fields transport energy and momentum from $q$ to $q_k$. Remember the different signs of the retarded and advanced Poynting\bz{-}vectors \eqref{ksdghnsdgsa}: Both the retarded and the advanced fields transport energy and momentum from the past to the future. Thus Cramer's simple model is able to describe radiation back\bz{-}reaction (which in this case might better be called recoil upon the source). 

While Cramer's work at best can be called a sketch for a theory to be developed in future, Hoyle and Narlikar\!\cite[parts III and IV]{Hoylenarl:actatdist} have already worked out a quantized action\bz{-}at\bz{-}a\bz{-}distance theory to appreciable detail. They claim to have achieved some remarkable advantages over \al conventional\ar  quantum field theory. 

\section{Conclusions}\label{absch:zusfass} 
The \raisebox{4\baselineskip}[0pt][0pt]{\hypertarget{ta:zusfass}{}}absorber theory of Wheeler and Feynman has been presented in it's non\bz{-}relativistic form. In particular we have emphasized: 
\begin{ggitemize}
\ggitem{Postulate \eqref{ofdnjgkmfc}, according to which retarded interactions must be handled strictly different from advanced interactions, can not be reconciled with classical action\bz{-}at\bz{-}a\bz{-}distance electrodynamics, because it violates the symmetry of this theory under time inversion. On the other hand, that postulate is indispensable for absorber theory; none of it's essential results can be achieved without that postulate.} 
\ggitem{The range of applicability of postulate \eqref{ofdnjgkmfc} is restricted to stationary, not time\bz{-}dependent radiation processes. In case of time\bz{-}dependent radiation processes, that postulate leads to results, which are incompatible with observation.} 
\ggitem{The claim of Wheeler and Feynman, that the fields nearby a source may be interpreted as the superposition of the advanced or retarded fields emitted by the source (contributing \mbox{50\,\%} to the total amplitude) and the advanced fields emitted by the absorber particles (contributing the other \mbox{50\,\%} to the total amplitude) has been refuted. We have pointed out that due to this flaw absorber theory in any case is not compatible with experimental observation, no matter whether the strange postulate \eqref{ofdnjgkmfc} is accepted or not.} 
\ggitem{\al Derivation IV\ar  is neither based on the perplexing postulate \eqref{ofdnjgkmfc} nor is it affected by the error mentioned in the previous topic. Derivation IV however does not lead to the explicit expression \eqref{ksdgvsgc} for radiation back\bz{-}reaction, but merely to Dirac's cryptic expression \eqref{isayhgdrfd}.} 
\end{ggitemize}
It may not be superfluous to emphasize, that no objections have been raised against action\bz{-}at\bz{-}a\bz{-}distance electrodynamics. Just the contrary: In the last section we have put forward some arguments supporting the conjecture, that a quantum\bz{-}theoretical formulation of absorber theory probably would be free of the flaws of the classical theory. Such theory however, being a quantum theory, could of course by definition not explain the classical formula \eqref{ksdgvsgc} of radiation back\bz{-}reaction. 

\renewcommand{\bibname}{References}
\flushleft{\bibliography{../../gg}} 
\end{document}